\documentclass[prx,twocolumn,secnumarabic,amssymb,showpacs, superscriptaddress]{revtex4-1}
\usepackage{longtable}

\newcommand{\be}{\begin{eqnarray}}
\newcommand{\ee}{\end{eqnarray}}

\newcommand{\bra}[1]{\langle{#1}|}
\newcommand{\ket}[1]{|{#1}\rangle}

\newcommand{\sz}[1]{\hat{\sigma}_{\mathrm{z},#1}}
\newcommand{\szz}{\hat{\sigma}_{\mathrm{z}}}
\newcommand{\sxx}[1]{\hat{\sigma}_{\mathrm{x},#1}}
\newcommand{\sy}[1]{\hat{\sigma}_{\mathrm{y},#1}}

\newcommand{\smm}[1]{\hat{\sigma}_{-,#1}}
\newcommand{\spp}[1]{\hat{\sigma}_{+,#1}}

\newcommand{\catp}{\ket{\mathcal{C}^+_\alpha}}
\newcommand{\catm}{\ket{\mathcal{C}^-_\alpha}}

\newcommand{\catpis}{\ket{\mathcal{C}^+_{i\alpha}}}
\newcommand{\catmis}{\ket{\mathcal{C}^-_{i\alpha}}}

\newcommand{\sx}{\hat{\sigma}_{\mathrm{x}}}
\newcommand{\sm}{\hat{\sigma}_{-}}

\newcommand{\catpma}{\ket{\mathcal{C}^\pm_\beta}}
\newcommand{\catmpa}{\ket{\mathcal{C}^\mp_\beta}}

\newcommand{\catpa}{\ket{\mathcal{C}^+_\beta}}
\newcommand{\catma}{\ket{\mathcal{C}^-_\beta}}
\newcommand{\catpba}{\bra{\mathcal{C}^+_\beta}}
\newcommand{\catmba}{\bra{\mathcal{C}^-_\beta}}

\newcommand{\E}{\mathcal{E}}

\newcommand{\hta}{\hat{a}}
\newcommand{\htb}{\hat{b}}

\usepackage{mathtools}
\usepackage{dsfont}
\usepackage{xcolor}

\begin{document}

\title{{Stabilized Cat in Driven Nonlinear Cavity: A Fault-Tolerant Error Syndrome Detector }}
\author{Shruti Puri}
\affiliation{Department of Physics, Yale University, New Haven, CT 06520, USA}
\affiliation{Yale Quantum Institute, Yale University, New Haven, Connecticut 06520, USA}
\author{Alexander Grimm}
\affiliation{Yale Quantum Institute, Yale University, New Haven, Connecticut 06520, USA}
\affiliation{Department of Applied Physics, Yale University, New Haven, CT 06511, USA}
\author{Philippe Campagne-Ibarcq}
\affiliation{Yale Quantum Institute, Yale University, New Haven, Connecticut 06520, USA}
\affiliation{Department of Applied Physics, Yale University, New Haven, CT 06511, USA}
\author{Alec Eickbusch}
\affiliation{Yale Quantum Institute, Yale University, New Haven, Connecticut 06520, USA}
\affiliation{Department of Applied Physics, Yale University, New Haven, CT 06511, USA}
\author{Kyungjoo Noh}
\affiliation{Department of Physics, Yale University, New Haven, CT 06520, USA}
\affiliation{Yale Quantum Institute, Yale University, New Haven, Connecticut 06520, USA}
\author{Gabrielle Roberts}
\affiliation{Department of Physics, Yale University, New Haven, CT 06520, USA}
\affiliation{Yale Quantum Institute, Yale University, New Haven, Connecticut 06520, USA}
\author{Liang Jiang}
\affiliation{Yale Quantum Institute, Yale University, New Haven, Connecticut 06520, USA}
\affiliation{Department of Applied Physics, Yale University, New Haven, CT 06511, USA}
\author{Mazyar Mirrahimi}
\affiliation{Yale Quantum Institute, Yale University, New Haven, Connecticut 06520, USA}
\affiliation{QUANTIC team, INRIA de Paris, 2 Rue Simone Iff, 75012, Paris, France}
\author{Michel H. Devoret}
\affiliation{Yale Quantum Institute, Yale University, New Haven, Connecticut 06520, USA}
\affiliation{Department of Applied Physics, Yale University, New Haven, CT 06511, USA}
\author{Steven M. Girvin}
\affiliation{Department of Physics, Yale University, New Haven, CT 06520, USA}
\affiliation{Yale Quantum Institute, Yale University, New Haven, Connecticut 06520, USA}

\date{\today}
\begin{abstract}
In quantum error correction, information is encoded in a high-dimensional system to protect it from the environment. A crucial step is to use natural, low-weight operations with an ancilla to extract information about errors without causing backaction on the encoded system. Essentially, ancilla errors must not propagate to the encoded system and induce errors beyond those which can be corrected. The current schemes for achieving this {\it fault-tolerance} to ancilla errors come at the cost of increased overhead requirements. An efficient way to extract error syndromes in a fault-tolerant manner is by using a single ancilla with strongly biased noise channel. Typically, however, required elementary operations can become challenging when the noise is extremely biased. We propose to overcome this shortcoming by using a bosonic-cat ancilla in a parametrically driven nonlinear cavity. Such a cat-qubit experiences only bit-flip noise and is stabilized against phase-flips. To highlight the flexibility of this approach, we illustrate the syndrome extraction process in a variety of codes such as qubit-based toric codes, bosonic cat- and Gottesman-Kitaev-Preskill (GKP) codes. Our results open a path for realizing hardware-efficient, fault-tolerant error syndrome extraction.
\end{abstract}

\maketitle
\section{Introduction}
To perform useful large-scale quantum computation, fragile quantum states must be protected from errors, which arise due to their inevitable interaction with the environment. To achieve this, strategies for quantum error correction (QEC) are continuously being developed. The key idea behind QEC is that natural errors and interactions generally involve low-weight operators. Therefore in order to protect quantum information, it is stored or encoded in a {\it logical qubit} using the non-local degrees of freedom of a high-dimensional system~\cite{preskill2013sufficient}. 
Here high-weight operators imply many-body operators, arising for example in a system of several qubits or operators involving many energy levels of a single high-dimensional physical system, arising for example in a harmonic oscillator. 
The high-weight operators characterizing the code-space of quantum information are called {\it stabilizers}
and are designed so that they commute with the logical qubit operators but anti-commute with the errors in the system~\cite{shor1995scheme,steane1996error,kitaev2003fault,gottesman1997stabilizer,nielsen2002quantum}. In the absence of errors, the system lies in the $+1$ eigenspace of the stabilizer and after an error occurs it moves to the $-1$ eigenspace. Consequently, the location and type of errors can be determined from the result of measuring the stabilizers, which are also known as an {\it error syndromes}. 
Measurement of these high-weight stabilizers would require engineering highly un-natural, many-body interactions in the system which is undesirable for practical implementation. A more reasonable approach is to synthesize stabilizer measurements via naturally available couplings with an {\it ancillary} system~\cite{shor1996fault}. However, interaction with the ancilla exposes the encoded system to more errors. In fact, if the measurement is not designed correctly, errors from the ancilla's noise channel can propagate to the encoded system and damage it beyond repair. Therefore for error correction to be successful such a catastrophic backaction must be eliminated.\\
\indent
To illustrate a typical approach for synthesizing stabilizer measurements, consider a system $\mathcal{M}$ (logical qubit) which encodes quantum information in $N$ subsystems (physical qubits) and let $\hat{S}$ be a stabilizer. A code is defined by multiple stabilizers but, for simplicity we just consider one. Let ${\hat{M}_i, i=1,2,..N}$ be a set of low-weight operators which commute with $\hat{S}$ and can be used to synthesize $\hat{S}$ through coupling with an ancilla. As an example, the four-qubit operator $\sz{1}\sz{2}\sz{3}\sz{4}$ is a stabilizer for surface codes~\cite{raussendorf2007fault}, in which case $\hat{M}_i=\sz{i}$. On the other hand, the stabilizer for single-mode bosonic cat codes is the parity operator $\hat{P}=\exp(i\pi\hta^\dag\hta)$, in which case $\hat{M}=\hta^\dag\hta$~\cite{leghtas2013hardware,mirrahimi2014dynamically,ofek2016extending}. Here $\szz$ is a Pauli operator, while $\hta,\hta^\dag$ are the photon annihilation and creation operators. The ancilla is typically a qubit which is coupled to the encoded system via the interaction Hamiltonian
\begin{align}
\hat{V}=\sx \sum_{i=1}^{N}g_i(t)\hat{M}_i,
\label{H0}
\end{align}
where $\sx$ is the Pauli operator of the ancilla qubit and $g_i$ are controllable interaction strengths.  The evolution of the system is described by the unitary, 
\begin{align}
\hat{U}(t)&=\mathcal{T}\exp\left(-i\int_0^t \hat{V}(\tau)d\tau\right)\\
&=\cos\left(\sum_{i=1}^{N}\int_0^tg_i(\tau)\hat{M}_id\tau\right)\nonumber\\
&+i\sin\left(\sum_{i=1}^{N}\int_0^tg_i(\tau)\hat{M}_id\tau\right)\sx.
\label{unit1}
\end{align}
The couplings and duration of evolution are chosen~\cite{nigg2013stabilizer,sun2014tracking} so that the above unitary (up to local rotations) at time $T$ becomes, 
\be
\hat{U}(T)=\frac{1+\hat{S}}{2}+\frac{1-\hat{S}}{2}\sx
\label{fst}
\ee
From Eq.~\eqref{fst} we see that the ancilla state undergoes a bit-flip at time $T$ conditioned on whether the stabilizer is $+1$ or $-1$. Thus, measurement of the ancilla yields the error syndrome. Remarkably, even though the starting Hamiltonian in Eq.~\eqref{H0} was low-weight, its unitary evolution involves high-weight operators. During the time interval $[0,T]$, the ancilla and the encoded system are entangled and it is crucial that errors in the ancilla do not propagate as uncorrectable errors in the encoded data. Achieving this property, also referred to as {\it fault-tolerance} is crucial for the success of QEC and would require that all possible errors in the ancilla commute with $\hat{U}(t)$ at all times. Note that the ancilla qubit's bit-flip error $\sx$ satisfies this condition. Therefore if a bit-flip occurs at any time $\tau$ during the interval $[0,T]$, then at time $T$ the state of the system is described by the unitary $\hat{U}(T-\tau)\sx\hat{U}(\tau)=\sx\hat{U}(T)=\sx(1+\hat{S})/2+(1-\hat{S})/2$. It is clear that the ancilla's bit-flip channel only introduces an error in measurement of the syndrome without causing any backaction on the encoded system. In this case, the fidelity of syndrome extraction can be recovered by simply repeating the protocol multiple times and taking a majority vote over the measurement outcomes. 
Importantly, note that dephasing $\szz$ and amplitude damping $\sm$ errors in the ancilla do not commute with $\hat{U}(t)$. In fact, a single $\szz$ error on the ancilla propagates as a high-weight error to the encoded system. \\
\indent
There are primarily three approaches for fault-tolerant extraction of error syndromes developed by Shor~\cite{shor1996fault}, Steane~\cite{steane1997active}, and Knill~\cite{knill2005scalable}. 
These methods are based on using several ancillas prepared in complex quantum states, several transversal (or bitwise) entangling gates between the data and ancilla qubits, followed by ancilla measurements. For example, in Shor's method, a single ancilla qubit is replaced with an $w$-qubit Greenberger-Horne-Zeilinger (GHZ) state, where $w$ is the weight of the stabilizer. Steane's approach requires a whole extra ancillary code block prepared in the encoded $\ket{0}_\mathrm{E}$ and $\ket{1}_\mathrm{E}$ states. Knill's method, based on error correction by teleportation, requires two ancillary code blocks prepared in the encoded Bell state $\ket{0}_\mathrm{E}\ket{0}_\mathrm{E}+\ket{1}_\mathrm{E}\ket{1}_\mathrm{E}$. Unfortunately, these approaches lead to rapidly growing overhead of computationally expensive entangling gates and ancilla hardware, which forces more stringent requirement on error rates and pushes large-scale fault-tolerant quantum computation further out-of-reach. 
Some error correcting codes, such as the surface code, are designed to be tolerant to certain amount of ancilla errors. However, the error correcting threshold significantly degrades in the presence of noisy ancillas~\cite{bravyi2013simulation}. Alternatively, efforts are being directed towards optimizing the ancilla hardware for achieving fault-tolerance~\cite{chao2017quantum,reichardt2018fault}. For example, recently a technique for syndrome extraction in bosonic-cat-codes based on three-level ancilla (or a qutrit) was demonstrated~\cite{rosenblum2018fault}. However, this technique only provides protection against first-order errors in the ancilla and is still susceptible to the second- and higher-order errors. Extending this scheme for higher-order protection would require additional drives, which may ultimately open up new sources of errors and back-propagation. An alternate technique for direct, fault-tolerant syndrome extraction in bosonic-cat-codes based on engineering a high-weight stabilizer Hamiltonian has been proposed~\cite{cohen2017degeneracy}. However, practical realization of this scheme is challenging and would require new experimental developments.

\begin{figure*}[ht]
 \centering
 \includegraphics[width=2\columnwidth]{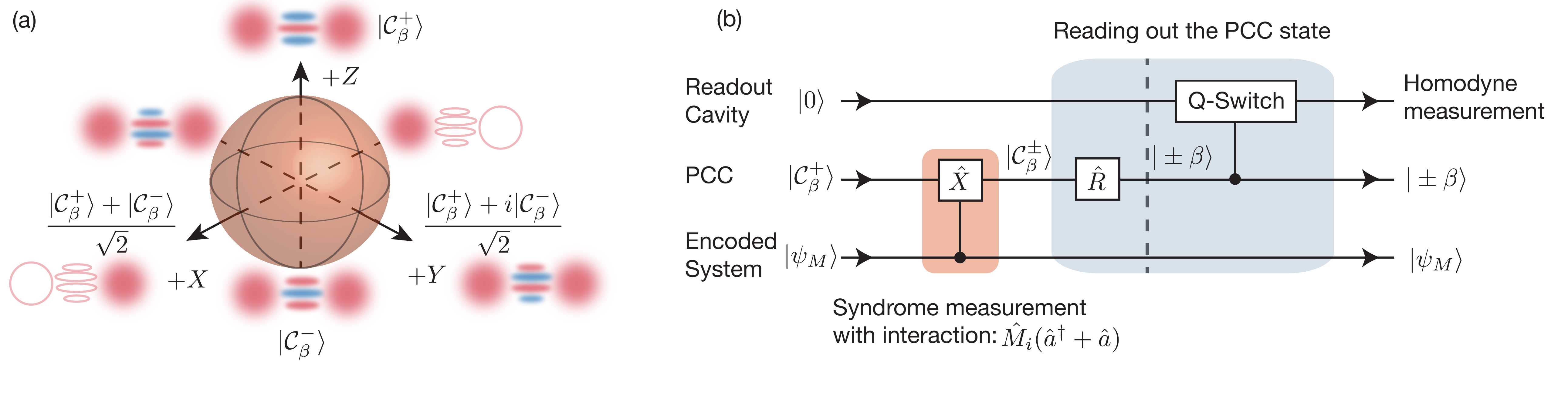}
 \caption{ (a) Bloch sphere representation of cat states. (b) The figure shows the overview of syndrome detection with a PCC. The first step is to map the error syndrome on the state of the PCC $\catpa$ or $\catma$ via a controlled ${X}$-rotation. The next step is to readout the state of the cat, which proceeds in two stages. The first stage is to rotate the cats $\catpma$ to the coherent states $\approx\ket{\pm\beta}$ (using the procedure described in the main text). In the second stage, a single-photon exchange coupling between the PCC and a low-Q readout cavity is turned on, a process known as {\it Q-Switch}. This coupling leads to displacement of the readout-cavity conditioned on the PCC state. Lastly, homodyne measurement of the signal from the low-Q cavity reveals the state of the PCC, thereby yielding the error syndrome.
}
 \label{overview}
 \end{figure*}

 \indent
The above discussion suggests that a more efficient way to extract the error syndrome fault-tolerantly is with an ancilla which exhibits a highly asymmetric error channel. In this case it would be possible to design a physical ancilla-storage unitary which would commute with the ancilla's error channel and will therefore be \textit{transparent} to ancilla errors~\cite{kapit2018error,sarlette2017loss}. Specifically if the ancilla only had $\sx$ errors, the example above shows that there would be a readout error on the syndrome but no backaction on the encoded system. With this in mind, we propose a two-component cat state in a pumped Kerr-nonlinear cavity as an ancilla for fault-tolerant syndrome extraction. The cat states $\catpma=\mathcal{N}^\pm_\beta(\ket{\beta}\pm\ket{-\beta})$ with $\mathcal{N}^\pm_\beta=1/\sqrt{2(1\pm e^{-2|\beta|^2})}$, are degenerate eigenstates of a Kerr-nonlinear cavity under two-photon driving~\cite{puri2017engineering,goto2016universal}. Note that these two states are orthogonal but have different normalization constants. We work in the basis so that the states along the $+Z$ and $-Z$ axis of the Bloch sphere, shown in Fig.~\ref {overview}(a), correspond to the cat states $\catpa$ and $\catma$ respectively. In this basis, the states along $+X$ and $-X$ axis correspond to the states $(\catpa\pm\catma)/\sqrt{2}$ which to an excellent approximation are the coherent states $\ket{\pm\beta}$ for large $\beta$. The remarkable property of such a pumped cat is that natural couplings can only cause rotations around the $X$ axis. Intuitively, this is because the pump creates a large energy barrier which prevents phase rotations (that is, rotation from the coherent states $\ket{\beta}$ to $\ket{-\beta}$ and vice versa). The error channel is dominated by bit-flip errors (which increase linearly with the size of the cat $|\beta|^2$ or equivalently the pump strength). But more importantly, the phase-flips and amplitude damping are exponentially suppressed (exponential in the size of the cat $|\beta|^2$ or the pump strength). As a result, this pumped cat-ancilla can be used for fault-tolerant syndrome measurements. Here, we will outline a general procedure to extract an error syndrome based on conditional rotation of the cat state around the $X$ axis using only low-weight local interactions. We will discuss the fault-tolerance of this technique in detail and examine specific examples based on three distinct error correcting codes, namely, qubit based toric codes~\cite{kitaev2003fault}, bosonic cat~\cite{leghtas2013hardware,mirrahimi2014dynamically} and Gottesman-Kitaev-Preskill (GKP) code~\cite{gottesman2001encoding}. Although these examples belong to the subclass of quantum codes known as stabilizer codes, the ideas for error syndrome extraction presented here could be extended to other types of codes as well. Finally we will show how the state of the cat ancilla can be read-out in an efficient manner. We find that the desired interactions between the encoded system and cat ancilla can be easily realized using the inherent nonlinearity of the ancilla itself. That is, no additional coupling elements are required. Our results are applicable in different quantum computing architectures and demonstrate the advantages of exploiting hardware-specific resources for achieving fault-tolerance in QEC.

\section{Pumped-cat syndrome detector}
Consider a Kerr-nonlinear cavity driven by a two-photon drive with frequency twice the frequency of the cavity. Its Hamiltonian in the rotating wave approximation is,
\be
\hat{H}_\mathrm{pcc}=-K\hta^{\dag 2}\hta^2+P(\hta^{\dag 2}+\hta^2).
\label{eq_pcc}
\ee
In the above expression, $\hta,\hta^\dag$ are the photon annihilation and creation operators, $K$ is the strength of the Kerr nonlinearity and $P$ is the strength of the two-photon drive. The cat states $\catpma$ or equivalently the coherent states $\ket{\pm\beta}$ are the degenerate eigenstates of this Hamiltonian where $\beta=\sqrt{P/K}$~\cite{puri2017engineering}. The two coherent states are quasi-orthogonal, $\langle\beta|-\beta\rangle=\exp(-2\beta^2)$, while the cat states are exactly orthogonal. Henceforth we will refer to this cavity as the pumped-cat cavity (PCC) and denote the cat subspace with $\mathcal{C}$. Note that, for $\beta=0$, $\ket{\mathcal{C}^+_0}=\ket{n=0}$ and $\ket{\mathcal{C}^-_0}=\ket{n=1}$, where $\ket{n=0}$ and $\ket{n=1}$ are the Fock states. The cat subspace is separated from the rest of the Hilbert-space $\mathcal{C}_\perp$ by an energy gap $\omega_\mathrm{gap}\propto 4K\beta^2$ (see Appendix A). $\omega_\mathrm{gap}$ is the gap in the frame which is rotating at the frequency of the cavity $\omega_\mathrm{pcc}$, while in the laboratory frame the energy gap is $\omega_\mathrm{pcc}-\omega_\mathrm{gap}$. The negative sign appears because the Kerr nonlinearity is attractive. Moreover the expression for this gap is only approximate and as $\beta$ approaches zero, the energy gap becomes $2K$ (which is also the gap between Fock states $\ket{n=0}$ or $\ket{n=1}$ and $\ket{n=2}$ in the rotating frame, see Appendix A). The PCC interacts with the encoded system $\mathcal{M}$ in such a way that the interaction Hamiltonian in the rotating frame is,
\begin{align}
\hat{H}_\mathrm{I}=\sum\chi_i(t)\hat{M}_i(\hta^\dag+\hta).
\label{intH}
\end{align}
In order to understand the effect of this coupling on the PCC, first note that the cat states undergo bit-flips under the action of the photon annihilation operator, $\hta\catpma=\beta p^{\pm 1}\catmpa$ where $p=\mathcal{N}^+_\beta/\mathcal{N}^-_\beta$ (see Appendix B). Recall that $\mathcal{N}^\pm_\beta=1/\sqrt{2(1\pm e^{-2\beta^2})}$ and for large enough $\beta$, $p\rightarrow 1$. While the action of the annihilation operator transforms a state within $\mathcal{C}$ to another state which also lies in $\mathcal{C}$, the photon creation operator $\hta^\dag$ can take the PCC out of $\mathcal{C}$. However, for small couplings $\chi(t)\langle\hat{M}_i\rangle$, these spurious out-of-subspace excitations are suppressed due to the energy gap between $\mathcal{C}$ and $\mathcal{C}_\perp$. In this restricted subspace $\mathcal{C}$, $\hta^\dag\catpma=\beta p^{\mp 1}\catmpa$ (see Appendix B) and Eq.~\eqref{intH} can very well be approximated as,
\begin{align}
\hat{H}_\mathrm{I}\equiv 2\beta{\hat{\tilde\sigma}}_x\sum\chi_i'(t)\hat{M}_i
\label{intH_apx}
\end{align}
Here $\chi'_i(t)=\chi_i(t) (p+p^{-1})/2\sim\chi_i(t)$ and ${\hat{\tilde\sigma}}_x=\catpa\catmba+\catma\catpba$ is the effective Pauli operator in $\mathcal{C}$. This entangling interaction is identical to Eq.~\eqref{H0} and leads to unitary evolution equivalent to Eq.~\eqref{unit1}. Again, the couplings $\chi_i$ and time are chosen so that the unitary evolution corresponding to Eq.~\eqref{intH_apx} at time $t=T$ is given by $\hat{U}(T)=(1+\hat{S})/2+(1-\hat{S}){\hat{\tilde\sigma}}_x/2$ (ignoring possible local rotations). As a result, the ancilla cat state in the PCC undergoes a bit-flip conditioned on the stabilizer being $\hat{S}=1$ or $\hat{S}=-1$. The error syndrome can be easily extracted by reading out the state of the cat at time $T$. Figure~\ref{overview}(b) provides an outline of the proposed syndrome extraction scheme and we will delve into details with specific examples shortly. Note that, in some cases it might be physically more convenient to implement a coupling like $\sum\chi_i(t)(\hat{L}_i\hta^\dag+\hat{L}^\dag_i\hta)$ where $\hat{L}^\dag_i+\hat{L}_i=\hat{M}_i$. It is possible to synthesize fault-tolerant stabilizer measurements with such interactions as well. In fact, we use such a coupling for syndrome extraction in GKP code in section IV.C. 

\begin{table*}[ht]
\centering
\begin{tabular}[c]{c| c| c}
\hline\hline
Noise type & Jump operator $\hat{O}$ in the restricted $\mathcal{C}$  & Jump operators \\
 & subspace of the PCC & as $\beta\rightarrow 0$\\
\hline
\hline
Single-photon loss  & $\beta\left[\left(\frac{p+p^{-1}}{2}\right)\hat{\tilde{\sigma}}_\mathrm{x}+i\left(\frac{p^{-1}-p}{2}\right)\hat{\tilde{\sigma}}_\mathrm{y}\right]$ & $\hat{\tilde{\sigma}}_-$\\
Single-photon gain  & $\beta\left[\left(\frac{p+p^{-1}}{2}\right)\hat{\tilde{\sigma}}_\mathrm{x}-i\left(\frac{p^{-1}-p}{2}\right)\hat{\tilde{\sigma}}_\mathrm{y}\right]$ & $\hat{\tilde{\sigma}}_+$\\
Pure dephasing  & $\beta^2\left[\left(\frac{p^2+p^{-2}}{2}\right)\hat{\tilde{\mathcal{I}}}-\left(\frac{p^{-2}-p^2}{2}\right)\hat{\tilde{\sigma}}_\mathrm{z}\right]$ & $\frac{1}{2}(\hat{\tilde{\mathcal{I}}}-\hat{\tilde{\sigma}}_\mathrm{z})$\\
Two-photon loss  & $\hat{\tilde{\mathcal{I}}}$ & $\hat{\tilde{\mathcal{I}}}$\\
\hline
\hline
\end{tabular}
\caption{ In general, interaction with the environment can lead to single-photon loss, single-photon gain, pure dephasing and two photon loss.  
When the coupling to the environment is smaller than the energy gap $\omega_\mathrm{gap}$ then excitations out of the cat subspace $\mathcal{C}$ are negligible and the dynamics of the PCC can be restricted in $\mathcal{C}$. In this effective two-level system, the effect of the noise source can be described with the Lindbladian $\mathcal{D}[\hat{O}]\hat{\rho}=\hat{O}\hat{\rho}\hat{O}^\dag-(\hat{O}^\dag\hat{O}\hat{\rho}+\hat{\rho}\hat{O}^\dag\hat{O})/2$ where $\hat{O}$ is the jump operator which depends on the type of noise. The Lindbladian is derived using the Born approximation, along with the assumption that the spectral density of the noise is flat around the PCC frequency $\omega_\mathrm{pcc}$. The noise spectral density at the gap frequency is assumed to be small. For more discussion on these approximations see Appendix D, E and F. The jump operators corresponding to single-photon loss, single-photon gain, pure dephasing and two photon loss in a PCC are listed here. Here $p=\sqrt{1-e^{-2\beta^2}}/\sqrt{1+e^{-2\beta^2}}$ and for large $\beta$, $(p+p^{-1})/2\sim 1$ while $(p^{-1}-p)/2\sim e^{-2\beta^2}$. Therefore we find that as the size of the cat state increases, the only effect of the environment is to cause bit-flips in the cat subspace. As $\beta$ approaches $0$, the cat states $\catpma$ approach the Fock states $\ket{n=0,1}$ respectively. In this limit, the effect of noise reduces to the jump operators in a conventional two-level system as listed in the third column here.  }
\label{table}
\end{table*}

\section{Single-photon loss} We now examine the noise channel of the PCC. The major source of noise in a cavity is single-photon loss, which arises from the single-photon exchange coupling with a bath.
From the previous discussion it is clear that if the coupling to the bath is smaller than the energy gap between the $\mathcal{C}$ and $\mathcal{C}_\perp$ subspaces, then the dynamics of the PCC is confined to the $\mathcal{C}$ subspace. In this restricted subspace the effective two-level master equation becomes (see Appendix C),
\begin{align}
\dot{\hat{\rho}}&=-i[\hat{H}_\mathrm{pcc},\hat{\rho}]+\kappa_\mathcal{C}\beta^2\mathcal{D}\left[p^{-1}\catpa\catmba+p\catma\catpba\right]\hat{\rho},
\label{me1_m}\\
&=-i[\hat{H}_\mathrm{pcc},\hat{\rho}]+\kappa_\mathcal{C}\beta^2\mathcal{D}\left[\frac{p+p^{-1}}{2}\hat{\tilde{\sigma}}_x+\frac{p^{-1}-p}{2}i\hat{\tilde{\sigma}}_y\right]\hat{\rho}\label{me1_2_m}
\end{align}
where  $\mathcal{D}[\hat{O}]\hat{\rho}=\hat{O}\hat{\rho}\hat{O}^\dag-\frac{1}{2}\hat{O}^\dag\hat{O}\hat{\rho}-\frac{1}{2}\hat{\rho}\hat{O}^\dag\hat{O}$. Here we assume that there are no thermal excitations in the bath, that is, the PCC can only lose photons but not gain them (see Appendix D). Note that, as long as the evolution is confined to the $\mathcal{C}$ subspace, Eq.~\eqref{me1_m} (or Eq.~\eqref{me1_2_m}) reduces to the common master equation of a cavity coupled to a bath, $\dot{\hat{\rho}}=-i[\hat{H}_\mathrm{pcc},\hat{\rho}]+\kappa_\mathcal{C}\mathcal{D}[\hat{a}]\hat{\rho}$ (because $\hta=p\beta\catma\catpba+p^{-1}\beta\catpa\catmba$). Appendix C presents numerical simulations which confirm the theoretically derived master equation above. It is evident from Eq.~\eqref{me1_m} that the single-photon exchange coupling with the bath leads primarily to a bit-flip error which is accompanied by an exponentially small phase-flip error $\propto (p^{-1}-p)\sim \exp(-2\beta^2)$. In other words Eq.~\eqref{me1_m} implies that the bath lifts the two-fold degeneracy of the $\mathcal{C}$ subspace by an amount exponentially small in the size $\beta^2$. Intuitively, this can be understood from the fact that the number of photons in the state $\catma$ and $\catpa$ differ by an exponentially small amount, $\catpba\hta^\dag\hta\catpa=\beta^2p^2$, $\catmba\hta^\dag\hta\catma=\beta^2/p^2$. It is more likely for a photon to be lost to the environment from $\catma$ than $\catpa$ and this asymmetry lifts the degeneracy between the states $\catpma$. However, since the difference in the photon numbers decreases exponentially with $\beta$, the states $\catpma$ are almost degenerate even for a moderately sized $\beta$, (such as $\beta\sim 2$, $\exp(-2\beta^2)=3.3\times 10^{-4}$). 

The preservation of the degenerate cat subspace makes the PCC an excellent meter for syndrome detection because coupling with the bath commutes with the interaction Hamiltonian Eq.~\eqref{intH_apx} and does not cause backaction on $\mathcal{M}$. Single photon loss to the bath does induce random flips between $\catpma$, which reduces the accuracy of the measurement. Nevertheless, since there is no backaction, the accuracy can be easily recovered by repeating the measurement a few times and taking a majority vote. In Appendices D, E, F, we examine in detail other sources of errors such as photon gain, pure-dephasing, two-photon loss and the results are summarized in table~\ref{table}. We find that, irrespective of the underlying source of noise, the PCC's error channel is reduced to bit-flip errors while the phase-flips are exponentially suppressed. It is also important to point out that it is quite possible that spurious excitations or sudden non-perturbative affects overcome the energy barrier and cause excitations to the $\mathcal{C}_\perp$ subspace. These excitations, although rare, can impede the fault-tolerance of syndrome measurements. However as we show in Appendix E, any dissipation such as single- or two-photon loss will autonomously correct for such leakage errors. Having shown that the cat manifold is stabilized against phase-flips, we now delve into the details of each stage of the syndrome extraction protocol. We begin by describing how the error syndrome of an encoded system is mapped on to the PCC (blue region in Fig.~\ref{overview}(b)) using specific examples.

\section{Specific examples for stabilizer measurements}

\subsection{Four-qubit stabilizer $\sz{1}\sz{2}\sz{3}\sz{4}$ in toric codes} The $n$-qubit $\hat{\sigma}_{\mathrm{z}}$ stabilizer arises frequently in the toric code which is a topological quantum error correcting code~\cite{kitaev2003fault}. Because of its significance in two-dimensional toric codes, here we will focus on the direct, eigen-space preserving measurement of the $\hat{S}_\mathrm{z}=\sz{1}\sz{2}\sz{3}\sz{4}$ stabilizer. The Hilbert space of $\hat{S}_\mathrm{z}$ is classified into even $\mathcal{E}$ and odd eigenspaces $\mathcal{O}$. The eight-fold degenerate even (odd) subspace comprises of states which are +1 (-1) eigenstates of $\hat{S}_\mathrm{z}$. We define $\mathcal{E}$ (even-subspace) and $\mathcal{O}$ (odd-subspace) to be the code and error subspace respectively, so that a measurement of $\hat{S}_\mathrm{z}$ will yield $-1$ or $1$, indicating if there was or was not an error. Direct measurement of $\hat{S}_\mathrm{z}$ would require a five-body interaction between the code qubits and an ancilla which is challenging to realize experimentally. Instead, we perform syndrome measurement with two-body interactions by replacing $\hat{M}_i$ with $\sz{i}$ in Eq.~\eqref{intH}. The resulting interaction Hamiltonian $\hat{H}_\mathrm{I}=\chi(t) \hat{S}'_z(\hta+\hta^\dag)$, where $\hat{S}'_\mathrm{z}=\sz{1}+\sz{2}+\sz{3}+\sz{4}$, has the form of a longitudinal qubit-cavity coupling and has been realized experimentally~\cite{touzard2018parametrically,touzardlongitudinal}. For simplicity we have assumed that all the interaction strengths are equal. Although it is possible to make the interaction strengths equal~\cite{touzard2018parametrically,touzardlongitudinal,rosenblum2018fault}, our scheme does not require them to be equal. As long as the interaction strengths are known, the duration of interaction with each qubit can be adjusted to perform the syndrome measurement. An alternate approach is to keep the duration of interaction fixed, but use a pair of bit-flip pulses for each qubit appropriately separated in time~\cite{nigg2013stabilizer}.
\begin{figure}
 \centering
 \includegraphics[width=\columnwidth]{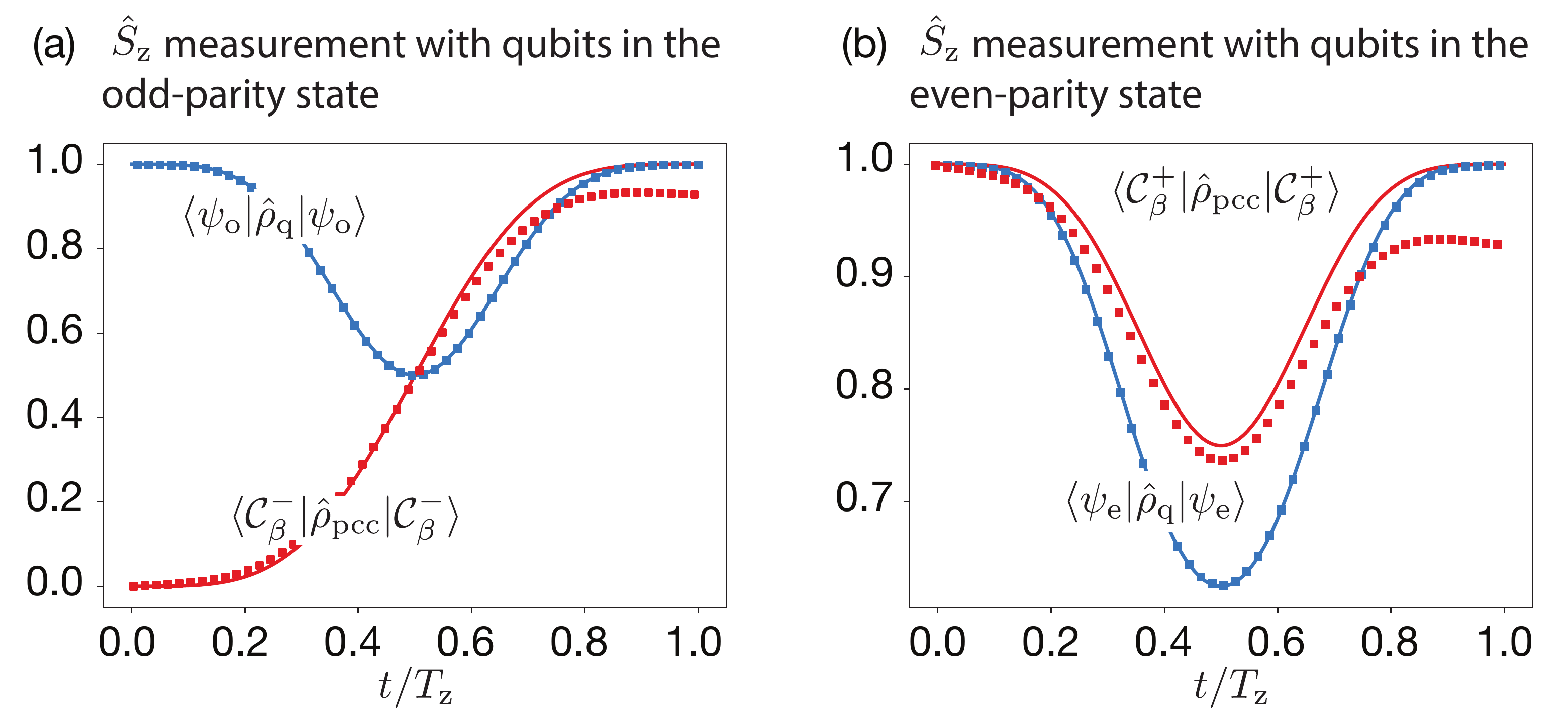}
 \caption{The figure shows the dynamics of the PCC and qubits during stabilizer measurement, when $\kappa=0$ (solid lines) and $\kappa=K/200$  (dotted lines). Here $\kappa$ is the rate of single-photon loss from the PCC. (a) Probability for the PCC and qubits to be in the state $\catma$ and $\ket{\psi_\mathrm{o}}$ when their initial states are $\catpa$ and $\ket{\psi_\mathrm{o}}$ respectively. (b) Probability for the PCC and qubits to be in the state $\catma$ and $\ket{\psi_\mathrm{e}}$ when their initial states are $\catpa$ and $\ket{\psi_\mathrm{e}}$ respectively. The states $\ket{\psi_\mathrm{o}}$ and $\ket{\psi_\mathrm{e}}$ are given in Eq.~\eqref{odd_q1} and Eq.~\eqref{even_q1}.  The parameters are $\chi=K/20$, $P=4K$ ($\beta=2$) and $T_\mathrm{z}=\pi/8\chi_0\beta$. Clearly when $\kappa=0$, the state of the PCC at time $T_\mathrm{z}$ reflects the syndrome $\langle \hat{S}_\mathrm{z}\rangle$. The probability for the PCC to correctly indicate the error syndrome is reduced to $\sim 93\%$ when $\kappa=K/200$ (red dotted lines). More importantly, as seen from the dotted blue lines, the state of the qubits after $T_\mathrm{z}$ is unaffected when $\kappa=K/200$.}
 \label{dyn1}
 \end{figure}
 \newline
 \indent
Following the analysis in section II, the unitary corresponding to this interaction becomes $\hat{U}(t)=i\sin\{2\beta\hat{S}'_\mathrm{z}\int_0^t\chi(\tau) d\tau\}{\hat{\tilde\sigma}}_\mathrm{x}+\cos\{2\beta\hat{S}'_\mathrm{z} \int_0^t\chi(\tau) d\tau\}$. To extract the syndrome, the PCC is initialized to the state $\catpa$ and the system evolved for a time $T_\mathrm{z}$ so that $\int_0^{T_\mathrm{z}}\chi(\tau)d\tau=\pi/8\beta$ (if the interaction strengths are unequal then the duration of interaction ${T_{i,\mathrm{z}}}$ must be such that $\int_0^{T_{i,\mathrm{z}}}\chi_i(\tau)d\tau=\pi/8\beta$). At this time the unitary reduces to, 
\be
\hat{U}(T_\mathrm{z})=e^{i\pi\hat{S}'_\mathrm{z}/4}\left[\left(\frac{1+\hat{S}_\mathrm{z}}{2}\right)+\left(\frac{1-\hat{S}_\mathrm{z}}{2}\right){\hat{\tilde\sigma}}_\mathrm{x}\right].
\label{unitc}
\ee
The first term in the above unitary ($\exp{(i\pi\hat{S}'_\mathrm{z}/4)}$) is just a local phase rotation of the qubits and can be kept track of in software while performing subsequent operations on qubits. Alternatively, local $\hat{\sigma}_\mathrm{z}$-gate can be applied to the qubit during or after syndrome measurement to compensate for these rotations.
It is clear that the state of the PCC after time $T_\mathrm{z}$ is $\catpa$ or $\catma$ if the qubits started in the code ($\hat{S}_\mathrm{z}=1$) or error subspace ($\hat{S}_\mathrm{z}=-1$). \\
\indent
We justify our theoretical analysis with exact numerical simulation of the master equation (ME) of the PCC and qubits in the presence of single-photon loss (for simplicity we assume the qubits to be lossless and use the common bosonic ME for the PCC),
\begin{align}
\dot{\hat{\rho}}&=-i[\hat{H},\hat{\rho}]+\kappa\mathcal{D}[\hta]\hat{\rho}, \\
H&=-K\hta^{\dag 2}\hta^2+P(\hta^{\dag 2}+\hta^2)+\chi(t) \hat{S}'_\mathrm{z}(\hta+\hta^\dag-2\beta).
\label{q_h}
\end{align}
Here $\kappa$ is the rate of single-photon loss of the PCC. All the numerical simulations in this work are carried out using a open source software~\cite{johansson2012qutip}. The last term in the above Hamiltonian ($-2\beta\chi(t)\hat{S}'_\mathrm{z}$) is added to cancel the deterministic single-qubit rotations (i.e., the term $\exp{(i\pi\hat{S}'_\mathrm{z}/4)}$ in Eq.~\eqref{unitc}). We take a time-dependent qubit-cavity interaction to simulate a realistic experimental setup where the coupling is switched on and then turned off. The qubits are initialized in a maximally entangled state in $\mathcal{O}$, $\ket{\psi_\mathrm{o}}$ 
\begin{align}
\ket{\psi_\mathrm{o}}=\frac{1}{\sqrt{8}}\left(\sum_i \hat{\sigma}_{\mathrm{x},i}+\sum_{i,j,k} \hat{\sigma}_{\mathrm{x},i}\hat{\sigma}_{\mathrm{x},j}\hat{\sigma}_{\mathrm{x},k}\right)\ket{0,0,0,0}.\label{odd_q1}
\end{align}
and the stabilized cat cavity is initialized to $\catpa$, with $P=4K$ ($\beta=2$), $\chi=(\pi/2)\chi_0\sin(\pi t/T_\mathrm{z})$, $\chi_0=K/20$ and $T_\mathrm{z}=\pi/(8\chi_0\beta)$. To begin with, the ME is solved with $\kappa=0$ to obtain the reduced density matrix of the PCC ($\hat{\rho}_\mathrm{pcc}$) and qubits ($\hat{\rho}_\mathrm{q}$). Figure~\ref{dyn1}(a) shows the probability for the PCC and qubits to be in the state $\catma$ (red) and $\ket{\psi_\mathrm{o}}$ (blue) respectively. As expected, after time $T_\mathrm{z}$ we find $\langle\psi_\mathrm{o}|\hat{\rho}_\mathrm{q}|\psi_o\rangle= 0.9999\sim 1$ and $\langle\mathcal{C}^-_\beta|\hat{\rho}_\mathrm{pcc}|\mathcal{C}^-_\beta\rangle= 0.9999\sim 1$. Next the effect of single-photon loss is studied by using $\kappa=K/200 (K/10)$. We find that at time $T_\mathrm{z}$, while the probability for the PCC to be in the $\catma$ state is reduced $\langle\mathcal{C}^-_\beta|\hat{\rho}_\mathrm{pcc}|\mathcal{C}^-_\beta\rangle=0.93 (0.52)$ because of loss-induced bit-flips between $\catpa$ and $\catma$, the qubits remain in the state $\ket{\psi_\mathrm{o}}$, $\langle\psi_\mathrm{o}|\hat{\rho}_\mathrm{q}|\psi_\mathrm{o}\rangle= 0.9999\sim 1$. We observe that although the fidelity of mapping the syndrome on to the ancilla cat is reduced to 52$\%$ for $\kappa=K/10$ (in which case the majority vote almost fails), there is still no backaction on the qubits. 

The analysis is repeated with the qubits and PCC initialized to $\ket{\psi_\mathrm{e}}$ and $\catpa$ respectively. Here,
\begin{align}
\ket{\psi_\mathrm{e}}=\frac{1}{\sqrt{8}}\left(\sum_{i,j} \hat{\sigma}_{\mathrm{x},i}\hat{\sigma}_{\mathrm{x},j}+\hat{I}+\hat{\sigma}_{\mathrm{x},1}\hat{\sigma}_{\mathrm{x},2}\hat{\sigma}_{\mathrm{x},3}\hat{\sigma}_{\mathrm{x},4}\right)\ket{0,0,0,0}.\label{even_q1}
\end{align}
As shown in Fig.~\ref{dyn1}(b) for $\kappa=0$ $\langle\psi_e|\hat{\rho}_\mathrm{q}|\psi_e\rangle\sim 1$ and $\langle\mathcal{C}^+_\beta|\hat{\rho}_\mathrm{pcc}|\mathcal{C}^+_\beta\rangle \sim 1$ at $t=T_\mathrm{z}$. Because of single-photon loss $\kappa=K/200(K/10)$ the probability to be in the state $\catpa$ decreases to 0.93 (0.52) but the state of the qubits is $\ket{\psi_\mathrm{e}}$ with probability $0.9999\sim 1$. Consequently, these numerical results confirm that the qubits are transparent to the errors in the PCC. The single-photon loss in the PCC reduces the fidelity of the syndrome extraction, but this can be recovered by repeating the protocol many times and taking a majority vote. For example, with $\kappa=K/200$ the fidelity of the controlled $\hat{X}$ rotation reduces to $0.93\%$ but by repeating the procedure $5$ times the probability of correctly mapping the syndrome to the PCC increases to $99.7\%$. In the example considered above, $\chi$ is small compared with the energy gap $\omega_\mathrm{gap}$. We note that, a large $\chi/\omega_\mathrm{gap}$ can cause phase-diffusion of the qubits and we study this effect in more detail in Appendix H. It is possible to extend the results in this section to measure the four-qubit stabilizer $\hat{S}_\mathrm{x}=\sxx{1}\sxx{2}\sxx{3}\sxx{4}$ with a single ancilla-cat based on Jaynes-Cummings type interaction between the PCC and the qubits (see Appendix G).

\subsection{Cat code stabilizer $e^{i\pi\hta^\dag_\mathrm{s}\hta_\mathrm{s}}$}

\begin{figure}
 \centering
 \includegraphics[width=\columnwidth]{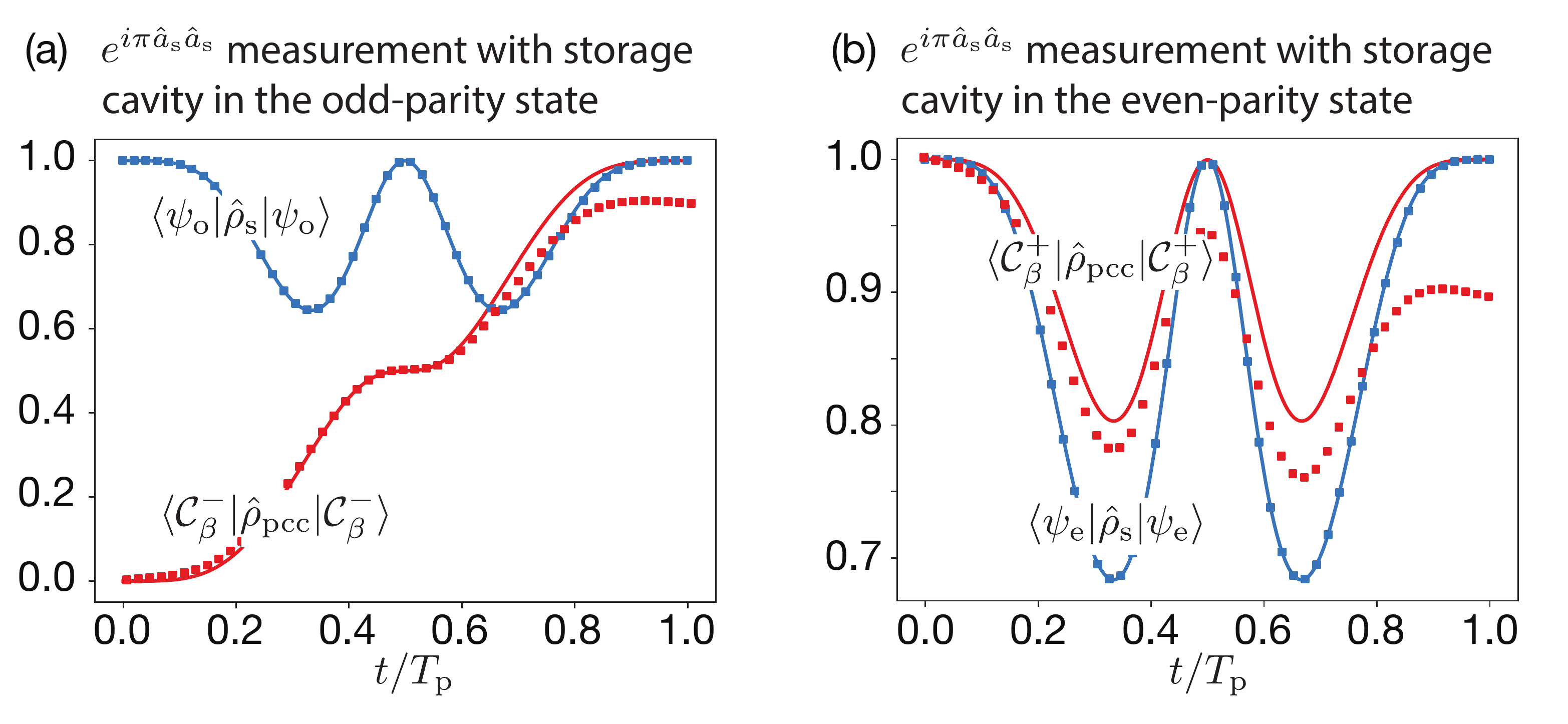}
 \caption{The figure shows the dynamics of the PCC and the storage-cat during stabilizer measurement when $\kappa=0$ (solid lines) and $\kappa=K/200$  (dotted lines). (a) Probability for the PCC and storage cavity to be in the state $\catma$ and $\ket{\psi_\mathrm{o}}$ when their initial states are $\catpa$ and $\ket{\psi_\mathrm{o}}$ respectively. (b) Probability for the PCC and the storage cavity to be in the state $\catma$ and $\ket{\psi_\mathrm{e}}$ when their initial states are $\catpa$ and $\ket{\psi_\mathrm{e}}$ respectively. Here $\ket{\psi_\mathrm{o}}=\catm+i\catmis$ and $\ket{\psi_\mathrm{e}}=\catp+\catpis$. The parameters are $\chi=K/15$, $P=4K$ ($\beta=2$), $\kappa=0$ and $T_\mathrm{p}=\pi/4\beta\chi_0$. Clearly the state of the PCC at time $T_\mathrm{p}$ reflects the photon-number-parity of the storage cat. The probability for the PCC to correctly indicate the error syndrome is reduced to $\sim 90\%$ when $\kappa=K/200$ (red dotted lines). However, as seen from the dotted blue lines, the state of the storage cat after $T_\mathrm{p}$ is unaffected by single-photon loss from the PCC.  }
 \label{dyn2}
 \end{figure}

Cat codes are bosonic error correcting codes where the information is encoded in superpositions of coherent states~\cite{leghtas2013hardware,mirrahimi2014dynamically}. The stabilizer for the cat code is the photon-number parity operator $\hat{P}=e^{i\pi\hta^\dag_\mathrm{s}\hta_\mathrm{s}}$ and indicates if there are even or odd number of photons. Here $\hta_\mathrm{s}$ and $\hta^\dag_\mathrm{s}$ are the photon annihilation and creation operators for the storage or data cat.
The two-fold degenerate code-subspace is defined by the cat states with even photon numbers: $\catp$ and $\catpis$, which are eigenstates of $\hat{P}$ with eigenvalue $+1$.  The error-space is comprised of the states with odd number of photons: $\catm$ and $\catmis$, which are eigenstates of $\hat{P}$ with eigenvalue $-1$. To avoid confusion we will refer to the cat states encoding quantum information as the storage cat. In the current scheme for cat syndrome measurement~\cite{sun2014tracking,ofek2016extending}, a storage cavity which encodes the cat codeword is coupled dispersively to an ancilla qubit. The dispersive coupling between the two is used to map the parity of the cat onto the ancilla. However, a random relaxation of the ancilla during the measurement induces a random phase rotation of the cat codeword, making this scheme non-fault tolerant~\cite{ofek2016extending,rosenblum2018fault}. In our approach, it is possible to achieve fault-tolerant syndrome detection by replacing the operator $\hat{M}$ with the photon number operator $\hat{n}=\hta^\dag_\mathrm{s}\hta_ \mathrm{s}$ in Eq.~\eqref{intH}. 
The interaction Hamiltonian of the storage cavity and PCC is given by $\hat{H}_\mathrm{I}=\chi(t) \hta^\dag_\mathrm{s}\hta_\mathrm{s}(\hta+\hta^\dag)$. This interaction, equivalent to a longitudinal interaction between the storage cavity and the PCC, can be realized in a tunable manner~\cite{touzard2018parametrically,touzardlongitudinal}. The unitary evolution generated by this interaction is
\begin{align}
\hat{U}(t)&=i\sin\left(2\beta\hta^\dag_\mathrm{s}\hta_\mathrm{s} \int_0^t\chi(\tau) d\tau\right){\hat{\tilde\sigma}}_\mathrm{x}\nonumber\\
&+\cos\left(2\beta\hta^\dag_\mathrm{s}\hta_\mathrm{s} \int_0^t\chi(\tau) d\tau\right).
\end{align}
The syndrome extraction proceeds by initializing the PCC to $\catpa$ and turning on the interaction between the storage-cavity and PCC for a time $T_\mathrm{p}$ so that $\int_0^{T_\mathrm{p}}\chi(\tau)d\tau=\pi/4\beta$. At this time, the unitary reduces to,
\begin{align}
\hat{U}(T_\mathrm{p})&=e^{-i\pi\hta^\dag_\mathrm{s}\hta_\mathrm{s}/2 }\left\{\left(\frac{1-\hat{P}}{2}\right){\hat{\tilde\sigma}}_\mathrm{x}+\left(\frac{1+\hat{P}}{2}\right)\right\}
\end{align}
The first term in the above equation $e^{(-i\pi\hta^\dag_\mathrm{s}\hta_\mathrm{s}/2 )}$ is just a deterministic rotation of the frame of reference of the storage cat which can be taken into account in software prior to further operations. If the storage is in the code subspace $x\catp+y\catpis$, then the state of the PCC and storage at time $T_\mathrm{p}$ is $\catpa$ and $x\catp+y\catpis$ respectively (up to a deterministic frame rotation of the storage cat). On the other hand, if the storage cat is in the error subspace $x\catm+y\catmis$, then the PCC evolves to the state $\catma$ at $T_\mathrm{p}$ while the storage cat remains in the state $x\catm+y\catmis$ (up to a deterministic frame rotation). Therefore the state of the cat in the PCC indicates the error syndrome $\hat{P}$. The PCC only measures the parity of the storage without revealing information about the actual photon statistics as long as $\chi$ is small and the dynamics of the PCC can be restricted to the stabilized cat manifold. For finite $\chi/K\beta^2$ there is a small probability of excitations out of the $\mathcal{C}$ subspace which could cause phase diffusion in the storage cat. 
Partial correction is possible by applying a counter-drive to the PCC to cancel the excitations out of the $\mathcal{C}$ subspace on average $\hat{H}_\mathrm{c}=- \chi \langle\hta^\dag_\mathrm{s}\hta_\mathrm{s}\rangle(\hta+\hta^\dag)$ (see Appendix I). 

The theoretical results are confirmed with numerical simulations of the master equation of the PCC and storage cavity in the presence of single-photon loss (for simplicity we assume the storage cavity to be lossless and use the common bosonic master equation for the PCC),
\begin{align}
\dot{\hat{\rho}}&=-i[\hat{H},\hat{\rho}]+\kappa\mathcal{D}[\hta]\hat{\rho},\\
 \hat{H}&=\hat{H}_\mathrm{pcc}+\chi(t) (\hta^\dag_\mathrm{s}\hta_\mathrm{s}-\langle\hta^\dag_\mathrm{s}\hta_\mathrm{s}\rangle)(\hta+\hta^\dag-2\beta).
 \label{catH}
\end{align}
The last term in the above Hamiltonian is added to compensate for the deterministic frame rotation of the storage cat $(e^{-i\pi\hta^\dag_\mathrm{s}\hta_\mathrm{s}/2 })$. The storage cavity is initialized in an odd-parity state $\ket{\psi_\mathrm{o}}=\catm+i\catmis$ and the stabilized cat cavity is initialized to $\catpa$, with $\alpha=2$, $P=4K$ ($\beta=2$), $\chi=(\pi/2)\chi_0\sin(\pi t/T_\mathrm{p})$, $\chi_0=K/15$ and $T_\mathrm{p}=\pi/(4\chi_0\beta)$. To begin with, $\kappa=0$ and the density matrix of the system is numerically estimated, from which the reduced density matrix of the PCC ($\hat{\rho}_\mathrm{pcc}$) and storage cavity ($\hat{\rho}_\mathrm{s}$) are obtained. Figure~\ref{dyn2}(a) shows the probability for the PCC and storage cavity to be in the state $\catma$ (red) and $\ket{\psi_\mathrm{o}}$ (blue) respectively. As expected, after time $T_\mathrm{p}$ we find $\langle\psi_\mathrm{o}|\hat{\rho}_\mathrm{q}|\psi_\mathrm{o}\rangle= 0.9999$ and $\langle\mathcal{C}^-_\beta|\hat{\rho}_\mathrm{pcc}|\mathcal{C}^-_\beta\rangle=0.9999 $. Next we study the effect of single-photon loss by using $\kappa=K/200$. We find that although at $T_\mathrm{p}$, the probability for the PCC to be in the $\catma$ state is reduced $\langle\mathcal{C}^-_\beta|\hat{\rho}_\mathrm{pcc}|\mathcal{C}^-_\beta\rangle=0.90$ because of loss-induced bit-flips between $\catpa$ and $\catma$, the storage cavity remains in the state $\ket{\psi_\mathrm{o}}$, $\langle\psi_\mathrm{o}|\hat{\rho}_\mathrm{q}|\psi_\mathrm{o}\rangle= 0.9999$. 

We repeat this analysis but with the qubits and PCC initialized to the even parity state $\ket{\psi_\mathrm{e}}=\catp+\catpis$ and $\catpa$ respectively. As shown in Fig.~\ref{dyn2}(b) for $\kappa=0$, $\langle\psi_\mathrm{e}|\hat{\rho}_\mathrm{s}|\psi_\mathrm{e}\rangle= 0.9999$ and $\langle\mathcal{C}^+_\beta|\hat{\rho}_\mathrm{pcc}|\mathcal{C}^+_\beta\rangle = 0.9999$ at $T_\mathrm{p}$. Because of single-photon loss $\kappa=K/200$ the probability to be in the state $\catma$ decreases to 0.90 but the state of the qubits is $\ket{\psi_\mathrm{e}}$ with probability $0.9999$. Note that in the example presented above, $\chi/\omega_\mathrm{gap}$ is small. In Appendix I, we study the effect of increasing $\chi/\omega_\mathrm{gap}$, in more detail. We also observe that the approach described here can be extended to measure the stabilizer of binomial~\cite{michael2016new} and pair cat code~\cite{albert2018multimode}. Moreover, the syndrome extraction technique can be adapted to perform a bias-preserving CNOT between two PCCs. Such a bias-preserving CNOT is unique to the system of stabilized cat qubits and promises to improve threshold requirements in quantum error correcting codes~\cite{puricnot}.

\subsection{Gottesman-Kitaev-Preskill (GKP) code stabilizers}

The GKP code is a bosonic error correcting code which is designed to correct random displacement errors in the phase space~\cite{gottesman2001encoding,albert2018performance}. The codewords are the simultaneous $+1$ eigenstates of the phase-space displacements $\hat{S}_\mathrm{q}=\exp{(2i\sqrt{\pi}\hat{q})}=D(i\sqrt{2\pi})$ and $\hat{S}_\mathrm{p}=\exp{(-2i\sqrt{\pi}\hat{p})}=D(\sqrt{2\pi})$. Here, $\hat{q}$ and $\hat{p}$ are the position and momentum operators, defined as $\hat{q}=(\hta^\dag_\mathrm{s}+\hta_\mathrm{s})/\sqrt{2}$ and $\hat{p}=i(\hta^\dag_\mathrm{s}-\hta_\mathrm{s})/\sqrt{2}$ respectively. Also, $D(i\sqrt{2\pi})$ and $D(\sqrt{2\pi})$ are the displacement operators, where $D(\alpha)=\exp(\alpha\hta^\dag_\mathrm{s}-\alpha^*\hta_\mathrm{s})$. The two ideal GKP codewords are uniform superpositions of eigenstates of $\hat{q}$ at even and odd integer multiples of $\sqrt{\pi}$ respectively. 
These states are a sum of an infinite number of infinitely squeezed states and are unphysical (non-normalizable) because of their unbounded number of photons. More realistic codewords can be realized by replacing the infinitely squeezed state $\ket{\hat{q}=0}$ with a squeezed Gaussian state and replacing the uniform superposition over these states by an overall envelope function, such as a Gaussian, a binomial etc~\cite{gottesman2001encoding, terhal2016encoding}. Recently, the GKP codewords have been realized in trapped-ion oscillators~\cite{fluhmann2018encoding}.
The GKP code provides protection against low-rate errors which can be expanded into small phase-space-displacements of the oscillator given by $\exp(-iu\hat{q})$ and $\exp(-iv\hat{p})$~\cite{albert2018performance, noh2018improved}. The displaced GKP states are also the eigenstates of the stabilizers $\hat{S}_\mathrm{q}, \hat{S}_\mathrm{p}$ with eigenvalue $e^{i2\sqrt{\pi}u}$ and 
$e^{2i\sqrt{\pi}v}$ respectively. A measurement of the stabilizers would yield the eigenvalues and hence uniquely determine the displacement errors $u,v$. This is possible only when $|u|,|v|<\sqrt{\pi}/2$, that is, when the displacement error is smaller than half the translational distance ($\sqrt{\pi}$) between the two codewords. \\
\indent
A simple approach to measure the eigenvalues $e^{2i\sqrt{\pi}u}$, $e^{2i\sqrt{\pi}v}$ of $\hat{S}_\mathrm{q}, \hat{S}_\mathrm{p}$ is based on an adaptive phase-estimation protocol (APE)~\cite{berry2009perform, berry2001optimal, terhal2016encoding}. This approach is based on repetitive application of displacements to the storage cavity which are conditioned on the state of the ancilla~\cite{terhal2016encoding}. In this section, we present a fault-tolerant protocol for the APE of the stabilizers for GKP code using a stabilized cat in a PCC. We will not go into the rigorous details of APE for GKP codes, which can be found in~\cite{terhal2016encoding}. Instead we will focus on implementing it with the stabilized cat ancilla. To achieve the controlled displacement required for APE, the storage cavity is coupled to the PCC via a tunable single-photon exchange (or a beam-splitter) interaction, $\hat{H}=\hat{H}_\mathrm{pcc}+(g(t)\hta^\dag\hta_\mathrm{s}+g^*(t)\hta\hta^\dag_\mathrm{s})$~\cite{gao2018programmable}. Such a tunable single-photon exchange coupling can be easily realized with the three- or four-wave mixing capability of the PCC and external drives of appropriate frequencies~\cite{flurin2015superconducting,pfaff2016schrodinger}. For small $|g|$ this Hamiltonian can be effectively written in the cat subspace as
\begin{eqnarray}
\hat{H}'&=\hat{H}_\mathrm{pcc}+\beta\left(\frac{p+p^{-1}}{2}\right)\left(g(t)\hta_\mathrm{s}+g^*(t)\hta^\dag_\mathrm{s}\right)\hat{\tilde{\sigma}}_\mathrm{x}\nonumber\\
&-i\beta\left(\frac{p-p^{-1}}{2}\right)\left(g(t)\hta_\mathrm{s}-g^*(t)\hta^\dag_\mathrm{s}\right)\hat{\tilde{\sigma}}_\mathrm{y}.
\label{gkp_full}
\end{eqnarray}
For large amplitude $\beta$, the second term becomes negligibly small and evolution under the above Hamiltonian implements a controlled displacement along the position or momentum quadrature depending on the phase chosen for the coupling $g$. In this limit, when the phase and amplitude of the coupling $g(t)$ is chosen so that $g^*(t)=g(t)=|g(t)|$ and $\beta\int_0^{T_1} |g(t)|dt=\sqrt{\pi/2}$, the unitary corresponding to the above Hamiltonian reduces to,
\begin{align}
\hat{U}_1(T_1)&=D\left(-i\sqrt{\frac{\pi}{2}}\right)\left\{\left(\frac{\hat{\tilde{\sigma}}_\mathrm{x}+1}{2}\right)D(i\sqrt{2\pi}) \right.\nonumber\\
&\left.+\left(\frac{1-\hat{\tilde{\sigma}}_\mathrm{x}}{2}\right)\right\},
\label{Ut1}
\end{align}
which is the conditional displacement of the cavity required for APE of $\hat{S}_\mathrm{q}$ (see Fig. 5 in~\cite{terhal2016encoding}). Similarly, when $g(t)=i|g(t)|$, $g^*(t)=-i|g(t)|$  and $\beta\int_0^{T_2} |g(t)|dt=\sqrt{\pi/2}$,
\begin{align}
\hat{U}_2(T_2)&=D\left(-\sqrt{\frac{\pi}{2}}\right)\left\{\left(\frac{\hat{\tilde{\sigma}}_\mathrm{x}+1}{2}\right)D(\sqrt{2\pi}) \right.\nonumber\\
&\left.+\left(\frac{1-\hat{\tilde{\sigma}}_\mathrm{x}}{2}\right)\right\}.
\label{ideal2}
\end{align}
This implements the required conditional displacements for APE of $\hat{S}_\mathrm{p}$. 

\begin{figure}
 \centering
 \includegraphics[width=\columnwidth]{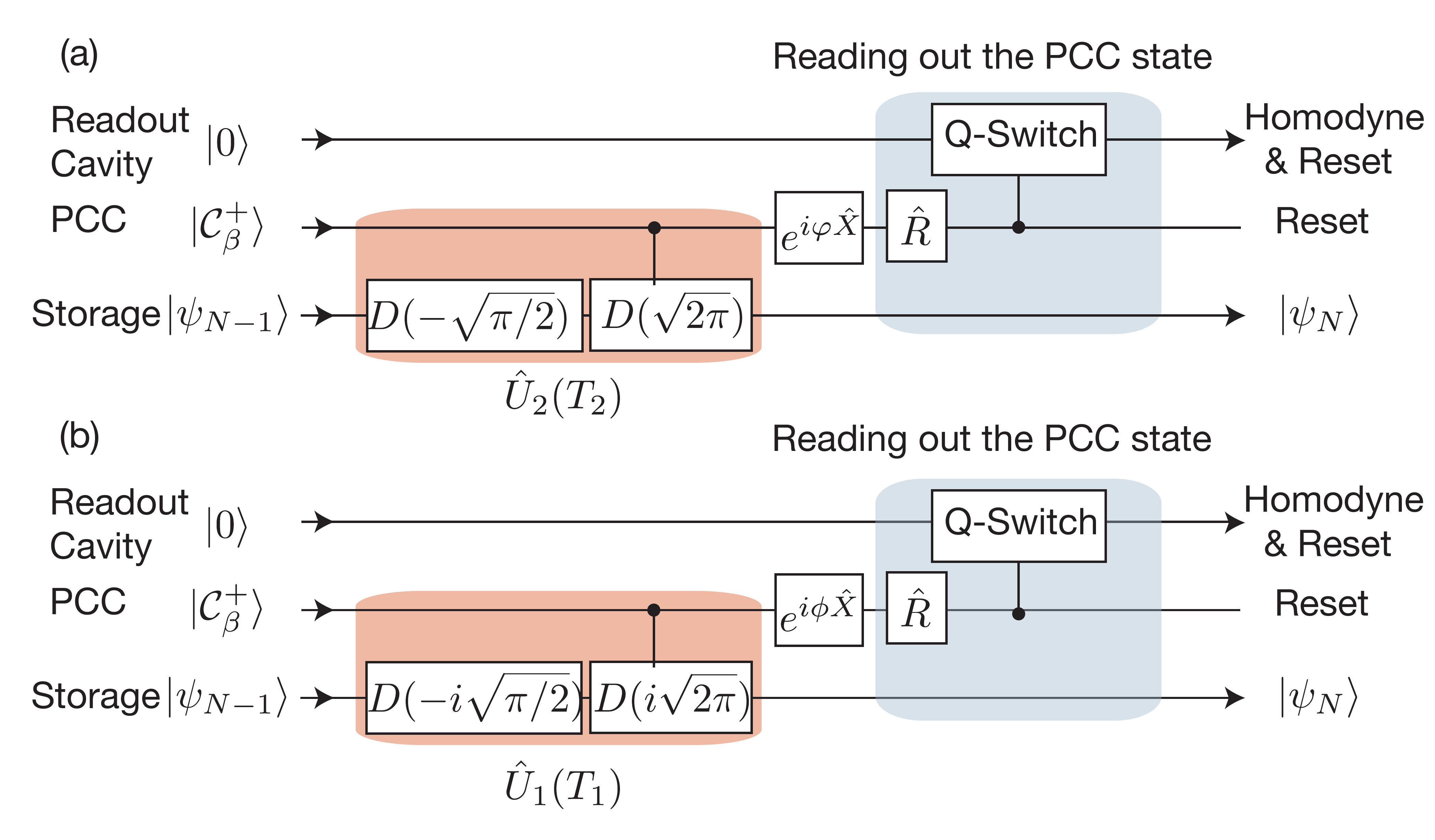}
 \caption{The figures (a,b) show the overall protocol of estimating the eigenvalues of $\hat{S}_\mathrm{p}$ and $\hat{S}_\mathrm{q}$ respectively with PCC. It proceeds by sequential application of the gates $\hat{U}_2(T_2)$ (or $\hat{U}_1(T_1)$) and measurement of the PCC. In APE, the state of the PCC is rotated by an angle $\phi$ or $\varphi$ around the $X$-axis of the Bloch sphere before each measurement. The angles $\phi$ and $\varphi$ are chosen based on its previous measurement record. }
 \label{gkp_pe1}
 \end{figure}

The overall protocol for APE is shown in Fig.~\ref{gkp_pe1}(a,b). For estimating  $\hat{S}_\mathrm{q}$, the protocol proceeds by sequential application of the gates $\hat{U}_1(T_1)$, followed by rotation of the PCC around the $X$-axis by an angle $\varphi$ and finally measurement of the PCC. Similarly, for estimating $\hat{S}_\mathrm{p}$, it proceeds by sequential application of the gates $\hat{U}_2(T_2)$, followed by rotation of the PCC around the $X$-axis by an angle $\phi$ and finally measurement of the PCC. The feedback phases $\phi$ and $\varphi$ are determined based on the measurement outcome in the previous round (Appendix J). As the number of rounds of phase estimation increases, the accuracy of the estimates for $u,v$ also increases and therefore, the uncertainty in the estimate of the eigenvalues $\exp(2i\sqrt{\pi}u)$ and $\exp(2i\sqrt{\pi}v)$ will decrease. The accuracy of the phase estimation protocol is evaluated using the Holevo phase variance $V_{\mathrm{q}}, V_{\mathrm{p}}$ which is defined as $V_{\mathrm{q},\mathrm{p}}=s_{\mathrm{q},\mathrm{p}}^{-2}-1$ with $s_{\mathrm{q}}=|\langle \hat{S}_\mathrm{q}\rangle|$ and $s_{\mathrm{p}}=|\langle\hat{S}_\mathrm{p}\rangle|$~\cite{terhal2016encoding}. For an ideal GKP state $V_{\mathrm{q},\mathrm{p}}=0$, while on the other hand, for large uncertainties in $u,v$ $V_{\mathrm{q},\mathrm{p}}\rightarrow\infty$. 

\begin{figure}
 \centering
 \includegraphics[width=.8\columnwidth]{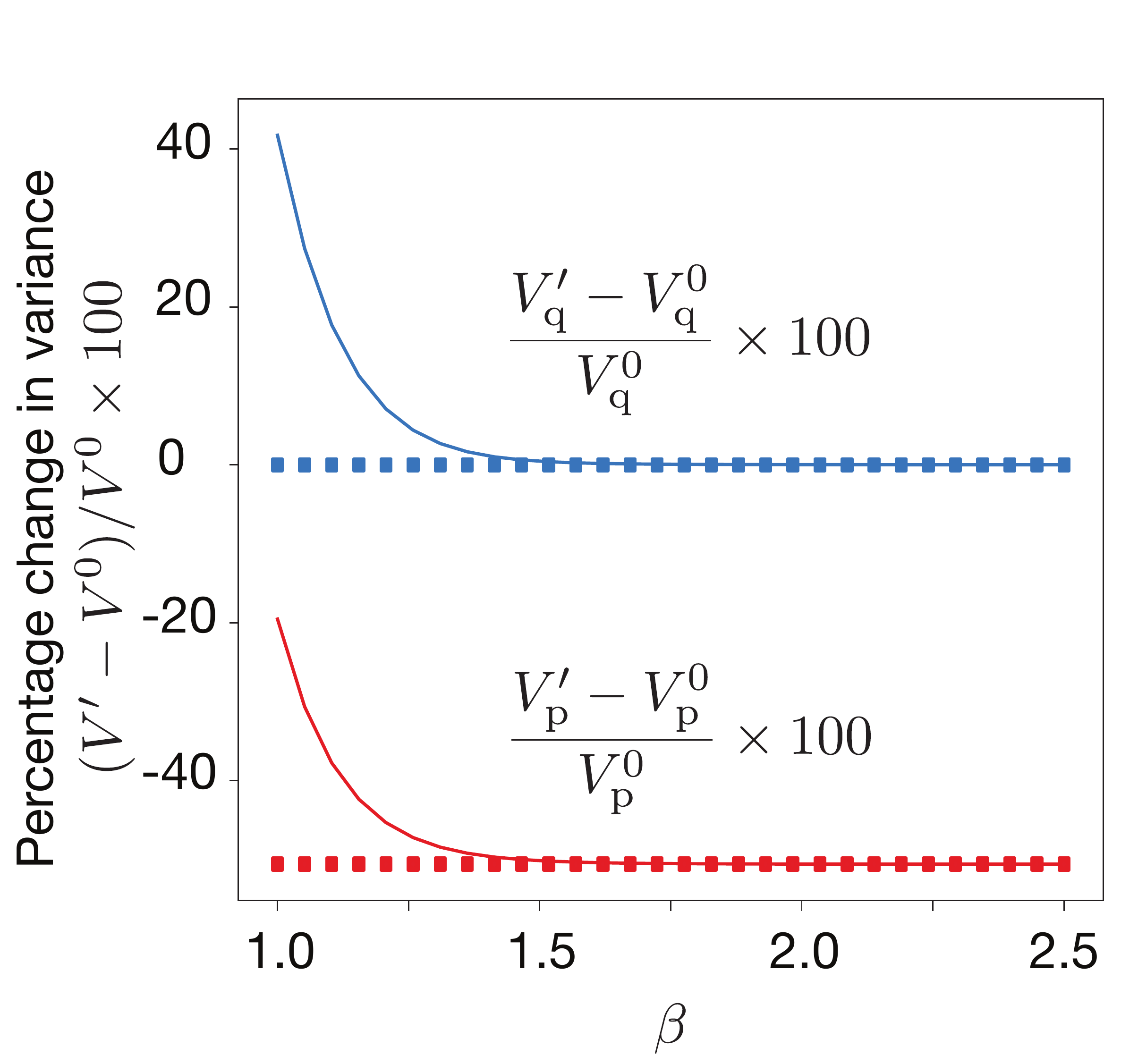}
 \caption{Percent change in the Holevo phase variances $V'_{\mathrm{q,p}}(T)$ for different sizes of the PCC cat state $\beta$ and fixed $g=0.02K$ for one round of phase estimation of $\hat{S}_\mathrm{p}$. The Kerr-nonlinearity of the PCC is fixed and the two-photon drive strength is varied as $P=K\beta^2$ so that the cat states with amplitude $\beta$ are stabilized. The solid lines show the change in variance in the $\hat{p}$ (red) and $\hat{q}$ (blue) quadratures. The dashed lines are the corresponding variances when phase estimation is carried out with an ideal two-level system. Since $\hat{S}_\mathrm{p}$ is being measured, the variance $V'_\mathrm{p}$ decreases (that is $V'_\mathrm{p}-V^0_\mathrm{p}<0$) while the variance $V'_\mathrm{q}$ remains unchanged (that is, $V'_\mathrm{q}-V^0_\mathrm{q}=0$). As expected, for large $\beta$ the solid and dotted lines converge and the APE protocol with the PCC becomes ideal. }
 \label{gkp_pe}
 \end{figure}

We numerically simulate one round of phase estimation for $\hat{S}_\mathrm{p}$ (that is, $g(t)=i|g(t)|$), with the storage in an approximate GKP state $\ket{\tilde{0}}_\mathrm{GKP}$,
\be
\ket{\tilde{0}}_\mathrm{GKP}&=N_0\sum_{n=-1}^1 \binom{2}{n+1}D\left(\sqrt{2\pi}n\right)\hat{S}_r\ket{0}.
\label{start_gkp}
\ee
In the above expression $N_0$ is the normalization coefficient, $\hat{S}_r=\exp\{r(\hta^2_s-\hta^{\dag 2}_s)/2\}$ is the squeezing operator 
with $r=1.4$ and the overlap function has been chosen to be the binomial coefficients $\binom{2}{n+1}$. The Holevo variance of this state is $V^0_\mathrm{q,p}=1.25,0.48$ (because we are starting with the approximate GKP state, $\ket{\tilde{0}}_\mathrm{GKP}$, the phase variance is not zero). The master equation used in the simulation is,
\begin{align}
\dot{\hat{\rho}}&=-i[\hat{H},\hat{\rho}]+\kappa\mathcal{D}[\hta]\hat{\rho},\\
\hat{H}&=\hat{H}_\mathrm{pcc}+ig(\hta^\dag_\mathrm{s}\hta-\hta_\mathrm{s}\hta^\dag).
\label{me_gkp}
\end{align}
The density matrices of the system is obtained at time $t=T=\sqrt{\pi}/(g\beta\sqrt{2})$. After this, the PCC is rotated around the $X$-axis by $\phi$, which will be taken to be $\pi/2$ (Appendix J).  After the projective measurement of the PCC states, the reduced density matrix for the storage cavity is obtained $\hat{\rho}_{s,\pm }$, from which the Holevo variance is evaluated along the $\hat{q}$ and $\hat{p}$ quadratures
\be 
V'_{\mathrm{q}}&=p_{+}V_{\mathrm{q}}(\hat{\rho}_{s,+ })+p_{-}V_{\mathrm{q}}(\hat{\rho}_{s,- }), \\
V'_{\mathrm{p}}&=p_{+}V_{\mathrm{p}}(\hat{\rho}_{s,+ })+p_{-}V_{\mathrm{p}}(\hat{\rho}_{s,- }).
\ee
A successful round of phase estimation for $\hat{S}_\mathrm{p}$ (or $\hat{S}_\mathrm{q}$) decreases the variance $V'_{\mathrm{p}}$ (or $V'_{\mathrm{q}}$). Figure~\ref{gkp_pe} shows the percent change in the variances, $100\times({V'}_\mathrm{q,p}-{V^0}_\mathrm{q,p})/{V^0}_\mathrm{q,p}$, for different sizes of the PCC cat state $\beta$ for $g=0.02K$ and $\kappa=0$. For comparison, we also simulate one round of ideal APE using a lossless two-level ancilla and estimate the resulting variances ${V'}^\mathrm{ideal}_\mathrm{q,p}$ (see Appendix L). The figure also shows the percent change in the variance $100\times({V'}^\mathrm{ideal}_\mathrm{q,p}-{V^0}_\mathrm{q,p})/{V^0}_\mathrm{q,p}$ (dashed line). Since $g$ is small, the dynamics of the PCC is confined within $\mathcal{C}$. For large $\beta$, as expected, phase estimation with PCC becomes increasingly accurate and the decrease in the Holevo variance is the same as with the ideal case. However, for small $\beta$ non-idealities due to the last term in Eq.~\eqref{gkp_full} are introduced and the magnitude of the decrease in variance becomes smaller.  

Let us now consider the effect of the photon-loss channel of the PCC. If the PCC undergoes a bit-flip during a round of phase estimation, the measurement outcome and hence the estimate for $u,v$ would be incorrect. This is equivalent to introduction of small displacement errors in the GKP state which can be corrected by repeated application of APE. More importantly, such errors do not increase the uncertainty in the phase variance. This can be confirmed by numerically solving the master equation in Eq.~\eqref{me_gkp} and evaluating the phase variance of the reduced density matrix of the storage cavity at time $T$, ${V}^\mathrm{m}_\mathrm{q,p}$ (see Appendix K). The variance calculated in this way corresponds to the situation when the measurement results after the APE are discarded. If, the observer (and environment) did not gain information about the system the variance ${V}^\mathrm{m}_\mathrm{q,p}$ should not change. Indeed, we find that as long as $\beta$ is moderately large (for example, $\beta=2$), then even for a large $\kappa$ (for example, $\kappa T=1$), ${V}^\mathrm{m}_\mathrm{q,p}-V^0_\mathrm{q,p}$ is negligible (for example $< 10^{-5}$). This shows that the interaction between the storage and PCC does not make the phase variance worse, which is the hallmark of fault-tolerance. Contrast this with the case when the phase estimation is carried out with a two-level system with relaxation noise rate $\gamma$ in time $T_\mathrm{ideal}$ (Appendix L). We find that for $\gamma T_\mathrm{ideal}=1$, ${V}^{\mathrm{m,ideal}}_\mathrm{p}-V^0_\mathrm{p}$ is negligible but ${V}^{\mathrm{m,ideal}}_\mathrm{q}-V^0_\mathrm{q}=+9.82$. The increase in the variance of the $\hat{q}$-quadrature signifies that relaxation actually made the phase variance (and hence the GKP state) worse. Clearly, unlike the PCC's error-channel, the storage is not transparent to the relaxation error of the two-level system.

\section{Reading out the ancilla cat} 
Once the error syndrome is mapped to the PCC, its state $\catpma$ must be determined. Although the readout of the PCC must be fast (so as to be able to repeat the protocol many times), it does not have to be QND, that is the readout can introduce phase-flips (or other errors) in the cat-ancilla. This is because the PCC-codeword interaction can be turned off while the PCC is being probed so that the ancilla errors don't propagate to the encoded system. Direct single-shot readout of cat states $\catpma$ is possible with another qubit. Such a high-fidelity ($>99\%$) readout has been demonstrated in superconducting circuits using transmons~\cite{sun2014tracking,ofek2016extending}. Here we discuss an alternate readout strategy which is based on measurement of the PCC along the $X$-axis of the Bloch sphere and does not require additional nonlinearities in the system. \\
\indent The states along the $X$-axis are (approximately) coherent states and can be measured easily using standard homodyne detection of the field at the output of the PCC. However the PCC is a (moderately) high-Q mode and so a direct homodyne measure will be slow. To overcome this we propose to {\it Q-switch} the PCC via frequency conversion into a low-Q readout cavity~\cite{flurin2015superconducting,pfaff2016schrodinger,gao2018programmable}. Because of the Q-switch, the low-Q readout cavity is displaced conditioned on the state of the PCC along the $X$-axis. Therefore a fast homodyne readout of the low-Q cavity reveals the state of the PCC and thereby the error syndrome. In the following we describe the process of rotation of cats from $\catpma$ to $\ket{\pm\beta}$ and the conditional displacement of the readout cavity. 
 
\subsection{Rotating the PCC cat states to coherent states}
To rotate the pumped cat $\catpma$, first a single-photon drive is applied so that its Hamiltonian is $\hat{H}=\E (\hta^\dag+\hta)-K\hta^{\dag 2}\hta^2+P(\hta^{\dag 2}+\hta^2)$. The single-photon drive rotates the pumped cat around the $X$-axis in time $T=\pi/8\E\beta$ from $\catpma$ to $(\catpa\pm i\catma)/\sqrt{2}$ (which is a {\it parityless cat}) respectively~\cite{mirrahimi2014dynamically,puri2017engineering, touzard2018coherent}. The state after time $T=\pi/8\E\beta$ is aligned along the $\pm Y$-axis of the Bloch sphere. 
A rotation around the $Z$-axis would then align the states along the $\pm X$-axis, which is however directly in contradiction with the fact that any natural interaction of the PCC only allows rotations around the $X$-axis. Therefore to achieve such an operation the two-photon pump is turned off and the states are allowed to evolve freely under the Kerr-nonlinear Hamiltonian $-K\hta^{\dag 2}\hta^2-K\hta^\dag\hta$ for a time $T=\pi/2K$ (the last term is added just for a phase reference). During this evolution the state $(\catpa\pm i\catma)/\sqrt{2}$ rotates to the (near) coherent state $(\catpa\mp\catma)/\sqrt{2}=\ket{\mp\beta}$~\cite{yurke1986generating,kirchmair2013observation,mirrahimi2014dynamically}. Next the two-photon pump is reapplied so that the cat subspace $\mathcal{C}$ is again stabilized against phase-flips. As a result, the PCC remains in the coherent states $\ket{\pm \beta}$ for a long time. Note that if there is a single-photon loss during the rotation around the $X$-axis then $\catpma$ can erroneously rotate to $(\catpa\mp i\catma)/\sqrt{2}$ respectively. On the other hand, while the two-photon drive is turned off, the single-photon loss can induce phase-errors (Appendix M). However, these errors only lead to a readout error and can be overcome by majority vote. 
\begin{figure*}[ht]
 \centering
 \includegraphics[width=2\columnwidth]{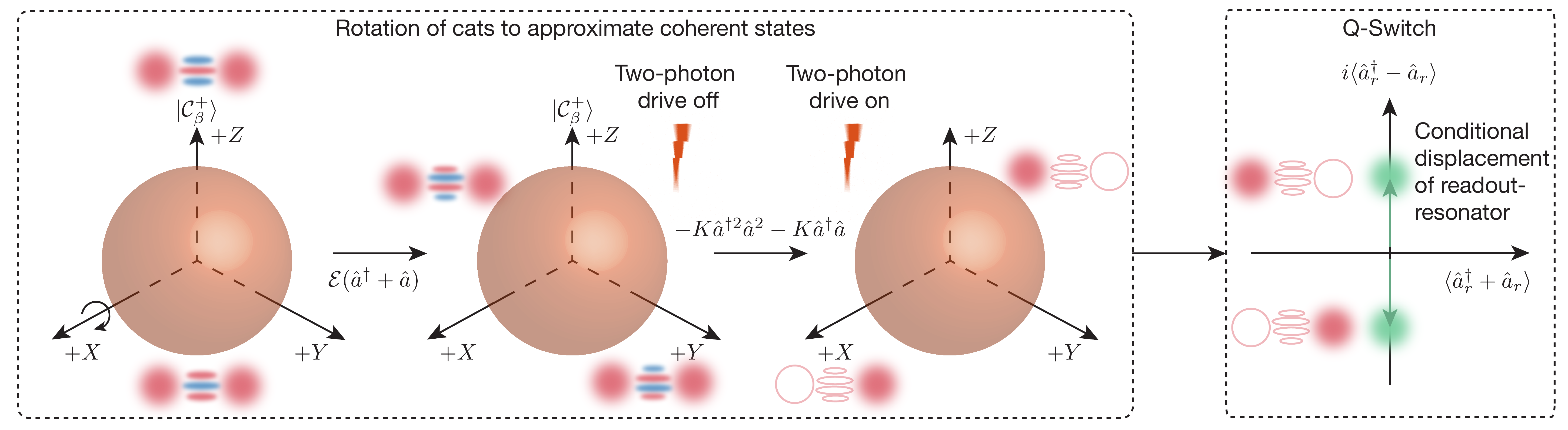}
 \caption{The figure shows an overview of the ancilla cat's readout cycle. After mapping the error syndrome, the state in the PCC is either $\catpa$ or $\catma$. These are first rotated around the $X$-axis to the states $(\catpa\pm i\catma)/\sqrt{2}$ respectively. The two-photon pump is then turned off and free evolution under the Kerr nonlinearity rotates $(\catpa\pm i\catma)/\sqrt{2}$ to $\sim\ket{\pm \beta}$ respectively. Following this rotation the two-photon drive is turned on such that the cat subspace $\mathcal{C}$ is again stabilized against rotations around the $Y$-axis or $Z$-axis. The next step is to switch-on the single-photon exchange coupling between the PCC and a low-Q readout cavity. This Q-switch operation displaces the readout cavity conditioned on if the PCC was in a coherent state $\ket{\beta}$ or $\ket{-\beta}$. Finally a homodyne detection of the field at the output of the readout reveals the state of the PCC, thereby extracting the error syndrome.  }
 \label{r_ov}
 \end{figure*}

\subsection{Q-Switching}
After the rotation described above, the state of the PCC lies along the $+X$ or $-X$-axis of the Bloch sphere (i.e., in the $\mathcal{C}$ manifold). The PCC is coupled to an off-resonant low-Q readout resonator (RR). In the absence of any external drives, the coupling between the two is negligible because of their large detuning. 
A single-photon exchange coupling (or a beam-splitter coupling) can be turned on by application of external drive(s) to compensate for the frequency difference between the PCC and the RR. A three- (or four) wave mixing between the drive(s), the PCC and RR results in a resonant single-photon exchange between the latter two. Such a controllable coupling has been implemented experimentally and is referred to as the Q-switch~\cite{flurin2015superconducting,pfaff2016schrodinger,gao2018programmable}. Once the Q-switch is turned on, the single-photon exchange coupling between the PCC and the readout cavity in the rotating frame is given by the Hamiltonian $\hat{H}_\mathrm{Q}=g(\hta^\dag\hta_\mathrm{r}+\hta\hta^\dag_\mathrm{r})$. For small $g$ this interaction can be re-written as 
\begin{align}
\hat{H}_\mathrm{Q}&=g\beta\left(\frac{p+p^{-1}}{2}\right)(\hta_\mathrm{r}+\hta^\dag_\mathrm{r})\hat{\tilde{\sigma}}_\mathrm{x}\nonumber\\
&-ig\beta\left(\frac{p-p^{-1}}{2}\right)(\hta_\mathrm{r}-\hta^\dag_\mathrm{r})\hat{\tilde{\sigma}}_\mathrm{y}.
\label{read_m}
\end{align}
Ignoring the term $\propto \hat{\tilde{\sigma}}_\mathrm{y}$, which becomes negligibly small even for moderately large $\beta$, the result of the Q-Switch is to displace the readout cavity conditioned on the state $(\catp\pm\catm)/\sqrt{2}$ of the PCC,
\be
\langle\hta_\mathrm{r}\rangle=\mp \frac{2 gi\beta}{\kappa_\mathrm{r}}(1-e^{-\kappa_\mathrm{r} t/2}).
\label{ideal}
\ee
In the above expression $\langle\hta_\mathrm{r}\rangle$ is the amplitude of the RR's field and $\kappa_\mathrm{r}$ is its linewidth. As a result, a homodyne detection of the readout cavity will measure the PCC in the $X$ basis and hence extract the error syndrome. At steady state $\langle\hta_\mathrm{r}\rangle=\langle\hta_\mathrm{r}\rangle_\mathrm{max}=2ig\beta/\kappa_\mathrm{r}$ and the measurement rate of the homodyne signal from the readout cavity is $R_\mathrm{ideal}=2\kappa_\mathrm{r}|\langle\hta_\mathrm{r}\rangle_\mathrm{max}|^2=8g^2\beta^2/\kappa_\mathrm{r}$. At the same time, for the PCC dynamics to be confined in $\mathcal{C}$ we require $g\langle\hta_\mathrm{r}\rangle_\mathrm{max}\ll 4K\beta^2$ which implies $(g^2/\kappa_\mathrm{r})\ll 2K\beta$. Therefore the measurement rate is limited by the energy gap between $\mathcal{C}$ and $\mathcal{C}_\perp$. Furthermore as $\mathrm{Im}[\langle\hta_\mathrm{r}\rangle]$ increases, the second term in Eq.~\eqref{read_m} can cause rotations around the $Y$ axis of the Bloch sphere thereby reducing $\hat{\tilde{\sigma}}_\mathrm{x}$ and the homodyne signal. However the rate of these rotations $= g\beta(p-p^{-1})\mathrm{Im}[\langle\hta_\mathrm{r}\rangle]/2$ is exponentially suppressed compared to the measurement rate even for moderately large $\beta$ (see Appendix N for numerical simulations of the Q-Switch operation).
  
\section{Discussion}
We have introduced a protocol to fault-tolerantly measure error syndromes, which is applicable for a variety of quantum error correcting codes such as qubit-based toric codes and various bosonic codes. The underlying principle of achieving fault-tolerance is to use a single ancilla with strongly asymmetric error channel. Preserving noise bias while being coupled to the relevant degrees of freedom of the encoded system is a demanding task. Even elementary operations, such as readout along the relevant axis can become challenging. However we show that the parametrically driven nonlinear cavity (PCC) is an excellent device to resolve the apparent incompatibility between noise bias and efficient control. \\
\indent
Another possible realization of a cat-qubit with strongly biased noise channel is based on engineering two-photon dissipation in a parametrically pumped cavity. The cat states are the steady states of this system and just like the PCC, small couplings with the environment only lead to bit-flips. The two-photon dissipation is realized by coupling the cat-cavity to another dissipative nonlinear element and applying drives at appropriate frequencies~\cite{mirrahimi2014dynamically,leghtas2015confining,touzard2018coherent}. Such a system has been implemented in superconducting circuits, however, a strong noise bias has not yet been observed~\cite{touzard2018coherent}. To achieve a strong noise bias any nonlinearity in the cavity and environmental couplings must be much smaller than the {\it Liouvillian gap} which depends on the strength of the engineered dissipation. However, in the realization described above, the cross-Kerr interaction between the dissipative-cat and the nonlinear element is larger than the Liouvillian gap. Heating in the nonlinear coupling element causes a large backaction on the dissipative-cats which leads to phase-flips, thereby making the noise channel unbiased. However, this is not a fundamental limitation and might be overcome with alternate realizations. In contrast, the PCC is in itself nonlinear and does not require an external nonlinear element for its implementation. Therefore, its cross-Kerr interaction with spurious modes in the system can be suppressed below the energy gap $\omega_\mathrm{gap}$ and the possibility of achieving a strong noise bias in this system is realistic. \\
\indent
Although the PCC can be realized in many quantum computing platforms, its implementation in superconducting circuits is especially promising. For example, the Josephson Parametric Amplifier (JPA), which is a widely used tool in superconducting circuits, realizes the Hamiltonian in Eq.~\eqref{eq_pcc}~\cite{wustmann2013parametric,puri2017engineering}. The PCC can also be implemented with a single junction or transmon embedded in a 3D cavity (in fact, the PCC is essentially a slightly anharmonic transmon). The nonlinearity of the junction/transmon gives rise to the fourth-order Kerr-nonlinearity. The two-photon drive can be realized by four-wave mixing with two microwave drives, one of which is red-detuned $\omega_\mathrm{pcc}-\delta$, while the other is blue detuned from the $\omega_\mathrm{pcc}+\delta$. All other couplings required for syndrome extraction can also be realized in a controllable manner via the four-wave mixing capability of the Kerr-nonlinearity. The remarkable property of the stabilization realized with the PCC is that it is fully controllable via the two-photon drive. Once the drive is turned off, the cavity can evolve freely under the Kerr-nonlinearity and rotate from cats to coherent states. This allows for subsequent readout of the PCC (and therefore extraction of the error syndrome) via Q-switching. To summarize, our results offer a realistic, hardware-efficient way for fault-tolerant error syndrome extraction in QEC.

\section{Appendix}
\subsection{Energy gap}
\begin{figure}
 \centering
 \includegraphics[width=.6\columnwidth]{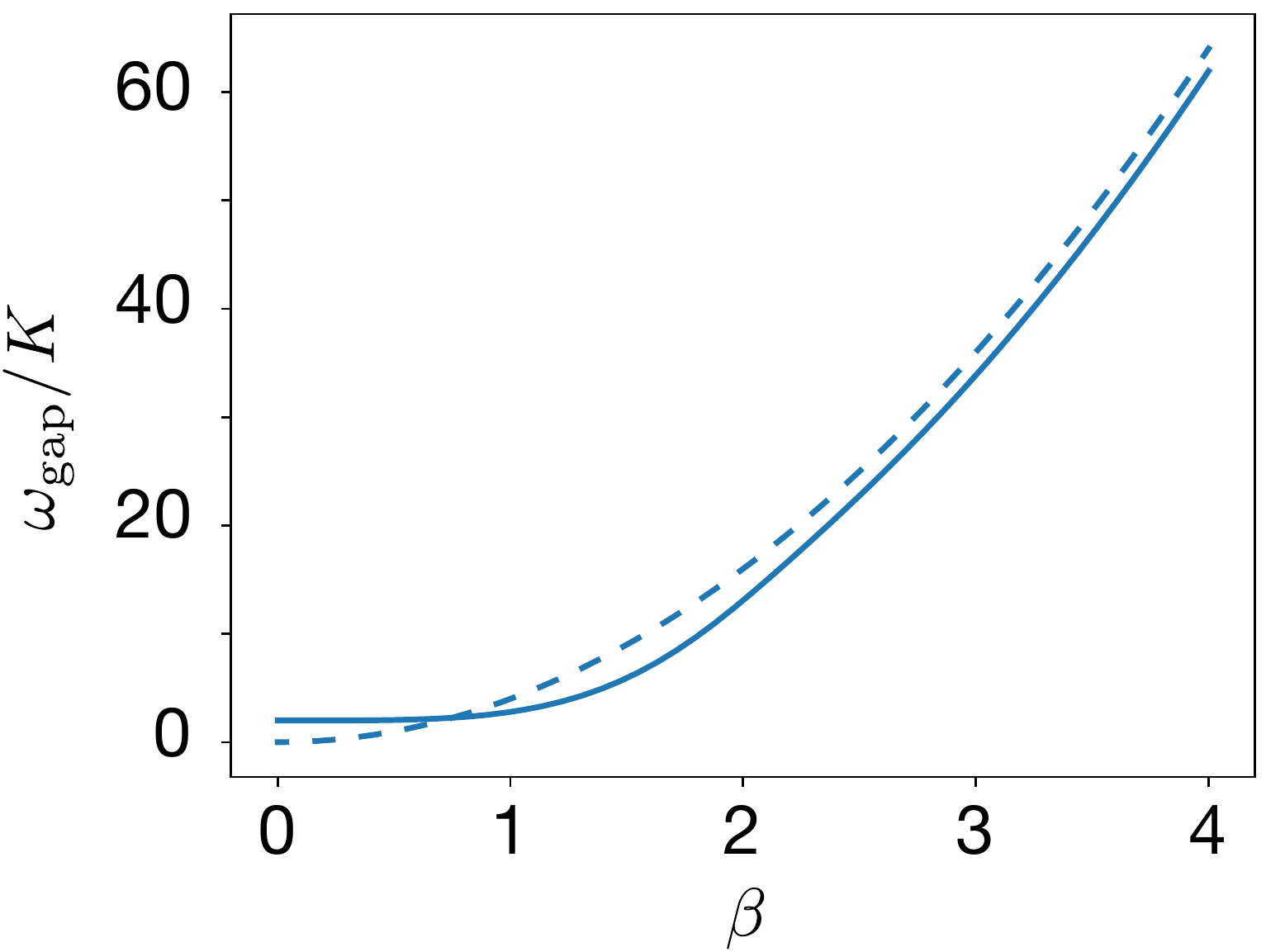}
 \caption{The figure shows the energy gap obtained by exact diagonalization of Eq.~\eqref{eq_pcc} (solid blue line) and the approximate gap $4K\beta^2$ (dashed blue line). As expected, the solid blue line converges to $2K$ as $\beta$ decreases. The approximate expression converges to the exact gap for large $\beta$, but it breaks down for small $\beta$ where the exact energy gap is $2K$. 
}
 \label{energy_gap}
 \end{figure}

Consider the Hamiltonian of the two-photon driven Kerr nonlinear cavity,
\be
\hat{H}=-K\hta^{\dag 2}\hta^2+P(\hta^{\dag 2}+\hta^2)
\label{eq_pcc}
\ee
The cat states $\catpma$ or equivalently the coherent states $\ket{\pm\beta}$ are the degenerate eigenstates of this Hamiltonian where $\beta=\sqrt{P/K}$~\cite{puri2017engineering}. We now make a displacement transformation
$D(\pm\beta)=\exp(\pm\beta\hta^\dag\mp\beta\hta)$ so that the above Hamiltonian reads
\begin{equation}
\begin{split}
\hat{H}'&  =-4K\beta^2\hta^\dag \hta-K\hta^{\dag 2}\hta^2\mp2K\beta (\hta^{\dag 2}\hta+\mathrm{h.c.}).
\end{split}
\end{equation}
In writing the above expression we have used $\beta=\sqrt{P/K}$ (so that the terms $\propto \hta^\dag,\hta,\hta^{\dag 2},$ and $\hta^2$ vanish) and also dropped the constant term $E=P^2/K$ which represents the energy of the coherent states $\ket{\pm\beta}$. The vacuum $\ket{0}$ is an eigenstate in this displaced frame (so that $D(\pm\beta)\ket{0}=\ket{\pm\beta}$ are the eigenstates in the original frame). In this frame, if we ignore the terms $\propto \beta^0,\beta^1$ (in the limit of large $\beta$) and consider only the term which is $\propto\beta^2$, then the next eigenstate is the Fock state $\ket{n=1}$. In the original frame, this would imply that the next eigenstates are $D(\pm\beta)\ket{1}$. The energy gap between $\ket{n=1}$ and $\ket{n=0}$ is $4K\beta^2$ and therefore this is also approximately the gap in the original frame. Figure~\ref{energy_gap} presents the energy gap as a function of $\beta$ evaluated by exactly diagonalizing Eq.~\eqref{eq_pcc} (solid blue line). It also shows the approximate gap $4K\beta^2$ (dashed blue line). Clearly, the approximate expression converges to the exact gap for large $\beta$. For small $\beta$ it becomes impossible to ignore the terms $\propto \beta^0,\beta^1$ and therefore the expression is incorrect. In fact if $\beta=0$, that is $P=0$, then Eq.~\eqref{eq_pcc} reduces to the Hamiltonian for an undriven nonlinear cavity and the energy gap becomes equal to that between Fock states $\ket{n=0}$ or $\ket{n=1}$ and $\ket{n=2}$, which is qual to $2K$.

\begin{figure}
 \centering
 \includegraphics[width=\columnwidth]{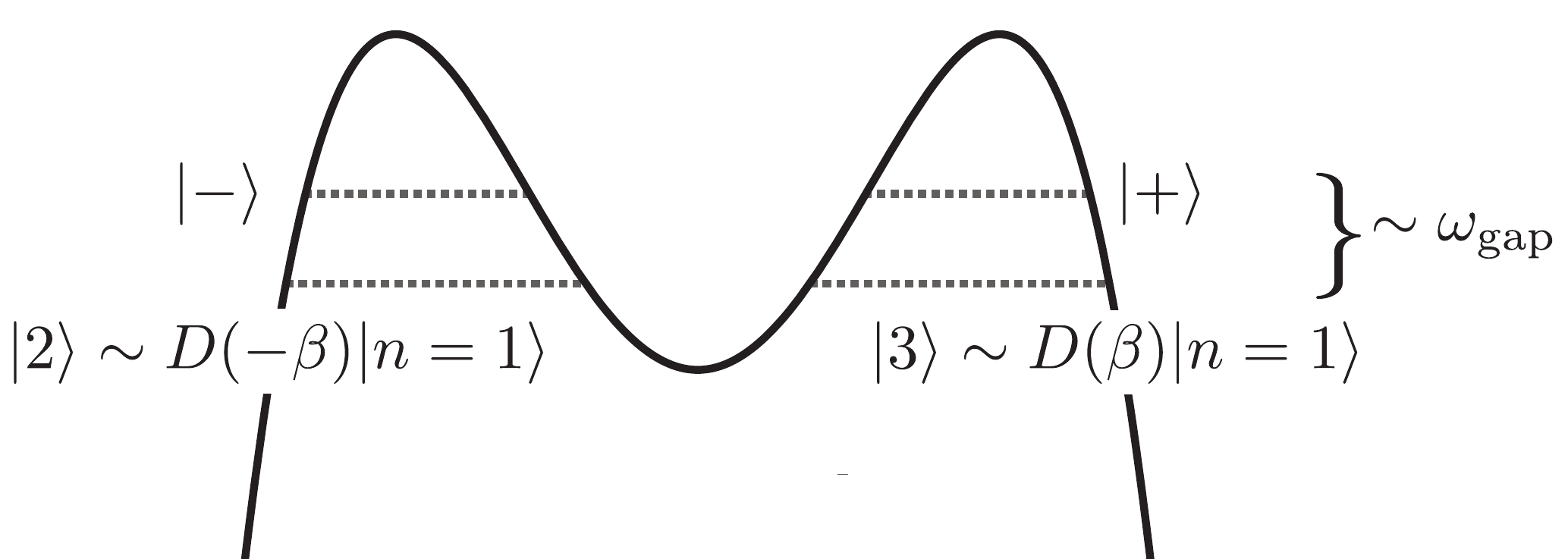}
 \caption{Illustration of the eigenspectrum of the PCC. The solid line represents the inverted-double well structure of the meta-potential corresponding to Eq.~\eqref{eq_pcc} while the dotted lines represents the energy levels (only four levels are shown). The cat states $\catpa$, $\catma$ or equivalently $\ket{+}$ and $\ket{-}$ are the exact degenerate eigenstates of the system ($\ket{\pm}=(\catpa\pm\catma)/\sqrt{2}$). As the pumping strength $P$ is increased, the meta-potential becomes deeper. In this case, the next two eigenstates are well approximated by the displaced Fock states $\ket{2}=D(-\beta)\ket{n=1}$ and $\ket{3}=D(-\beta)\ket{n=1}$ and are also approximately degenerate.   }
 \label{cartoon1}
 \end{figure}

At this point it will be useful to intuitively examine the eigenspectrum of Eq.~\eqref{eq_pcc}. The classical-potential or meta-potential of the PCC is an inverted double well with the states $\ket{+}\sim|\beta\rangle$ and $\ket{-}\sim \ket{-\beta}$ as two degenerate states, as shown in Fig.~\ref{cartoon1}~\cite{puri2017quantum}. The meta-potential is found by replacing the operators $\hta,\hta^\dag$ with complex numbers representing position and momentum~\cite{dykman2012fluctuating}. Note that the meta-potential does not show a physical energy landscape but gives a phase-space representation of a Hamiltonian. As we have already seen, $\catpa$, $\catma$ and equivalently their superposition states $\ket{+}$ and $\ket{-}$ are the exact degenerate eigenstates of the system, where $\ket{\pm}=(\catpa\pm\catma)/\sqrt{2}$. As the strength of the pump $P$ is increased, the two wells become deeper are pulled further apart. If the potential is deep enough, that is when $\omega_\mathrm{gap}\sim 4K\beta^2=4P$ is large, then the next two eigenstates are approximately degenerate as well and can be approximated by the displaced Fock states $\ket{2}=D(-\beta)\ket{n=1}$ and $\ket{3}=D(-\beta)\ket{n=1}$. This intuitive eigenspectrum will be useful in understanding the origin of phase-diffusion during stabilizer measurements in Appendix H and I.

\subsection{Photon annihilation and creation operators in the cat-subspace} Since coherent states are eigenstates of the photon annihilation operator, it is trivial to see that  $\hta\catpma=\beta p^{\pm 1}\catmpa$ where $p=\mathcal{N}^+_\beta/\mathcal{N}^-_\beta$. Therefore in the $\mathcal{C}$ subspace $\hta_\mathcal{C}=p\beta\catma\catpba+p^{-1}\beta\catpa\catmba$. Here $\hta_\mathcal{C}$ is the projection of $\hta$ in $\mathcal{C}$. Coherent states are not eigenstates of the photon creation operator. In fact $\hta^\dag\ket{\beta}=\beta\ket{\beta}+D(\beta)\ket{1}$ and $\hta^\dag\ket{-\beta}=-\beta\ket{-\beta}+D(-\beta)\ket{1}$, where $\ket{1}$ is the Fock state with one photon. In writing these expressions we are assuming $\beta$ is real for convenience. This implies that action of $\hta^\dag$ on coherent states $\ket{\pm\beta}$ or cat states $\catpma$ can cause leakage out of the code space. However if the energy gap is large, this leakage is suppressed. Therefore, the dynamics can be restricted to $\mathcal{C}$, in which $\hta^\dag_\mathcal{C}=p^{-1}\beta\catma\catpba+p\beta\catpa\catmba$. Here $\hta^\dag_\mathcal{C}$ is the projection of $\hta^\dag$ in $\mathcal{C}$. In other words, if the PCC is subject to a perturbation Hamiltonian which is expressed in terms of photon annihilation and creation operators, then as long as the Hamiltonian strength is smaller than the gap, the annihilation/creation operators can be replaced with their associated projections in the cat subspace ($\hta_\mathcal{C}$, $\hta^\dag_\mathcal{C}$).

\subsection{Master equation with single-photon loss}
The major source of noise in a cavity is single-photon loss, which arises from the single-photon exchange coupling with a bath $
\hat{H}_\mathrm{pcc,b}=\sum_k g_k(\hta\htb^\dag_k e^{i(\omega_k-\omega_\mathrm{pcc})t}+\hta^\dag\htb_ke^{-i(\omega_k-\omega_\mathrm{pcc})t})$. In this equation, $\hat{b}_k$ are the bath modes with frequency $\omega_k$ and $\omega_\mathrm{pcc}$ is the frequency of the PCC.
From the previous discussion it is clear that if the coupling to the bath is smaller than the energy gap $\omega_\mathrm{gap}$ between the $\mathcal{C}$, $\mathcal{C}_\perp$ subspaces and if there are no thermal excitations in the bath to compensate for this energy gap, then the dynamics of the PCC is confined to the $\mathcal{C}$ subspace. In this restricted subspace, the coupling between the PCC and bath becomes, 
\begin{align}
\hat{H}_\mathrm{pcc,b}&=\beta \sum_k g_k \left(p^{-1}\catpa\catmba+p\catma\catpba\right)\htb^\dag_k e^{i(\omega_k-\omega_\mathrm{pcc})t}\nonumber\\
&+\beta \sum_kg_k\left(p^{-1}\catma\catpba+p\catpa\catmba\right)\htb_k e^{-i(\omega_k-\omega_\mathrm{pcc})t}\label{me_r1}
\end{align}
The effective two-level master equation corresponding to the system-bath coupling described above can be derived as~\cite{carmichael2009statistical},
\begin{align}
\dot{\hat{\rho}}&=-i[\hat{H}_\mathrm{pcc},\hat{\rho}]+\kappa_\mathcal{C}\beta^2\mathcal{D}\left[p^{-1}\catpa\catmba+p\catma\catpba\right]\hat{\rho},
\label{me1}\\
&=-i[\hat{H}_\mathrm{pcc},\hat{\rho}]+\kappa_\mathcal{C}\beta^2\mathcal{D}\left[\frac{p+p^{-1}}{2}\hat{\tilde{\sigma}}_x+\frac{p^{-1}-p}{2}i\hat{\tilde{\sigma}}_y\right]\hat{\rho}\label{me1_2}
\end{align}
where  $\mathcal{D}[\hat{O}]\hat{\rho}=\hat{O}\hat{\rho}\hat{O}^\dag-\frac{1}{2}\hat{O}^\dag\hat{O}\hat{\rho}-\frac{1}{2}\hat{\rho}\hat{O}^\dag\hat{O}$. In deriving this master equation we assumed a flat-spectral density (Markov approximation) around $\omega_\mathrm{pcc}$. 

\begin{figure*}[ht]
 \centering
 \includegraphics[width=2\columnwidth]{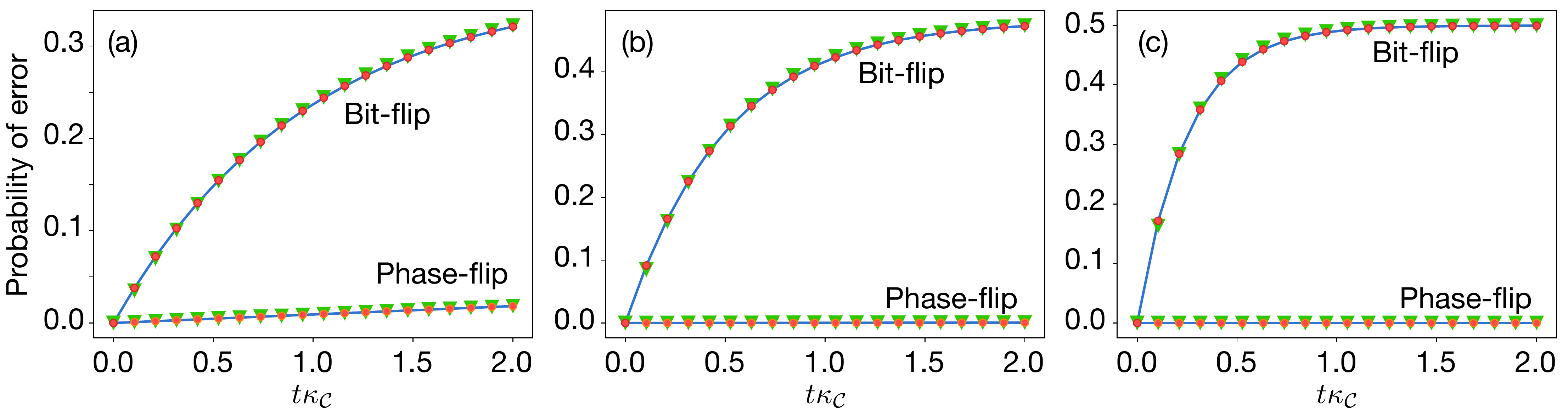}
 \caption{Comparison of dissipation predicted by (i) the effective two-level master equation (solid line) (ii) the common bosonic master equation (dots) and (iii) the master equation of a PCC coupled with a finite-linewidth cavity to emulate a general non-Markovian bath (triangles). In all the cases $K$ is fixed and $P$ is varied so that $\beta=1$, $\beta=\sqrt{2}$ and $\beta=2$ in panels (a), (b) and (c) respectively. For the dots $\kappa_\mathcal{C}=K/100$ and for 
the triangles $g=0.05K$, $\kappa_{bc}=2K$. The parameters are chosen such that $\kappa'_\mathcal{C}=\kappa_\mathcal{C}$. The probability of bit-flip error is estimated by initializing the PCC to the state $\catpa$ and estimating $\catmba\hat{\rho}_\mathrm{pcc}\catma$. The probability of phase-flip error is estimated by initializing the PCC to the state $\ket{+}$ and estimating $\bra{-}\hat{\rho}_\mathrm{pcc}\ket{-}$. 
}
 \label{dissipation_comp}
 \end{figure*}

We now provide numerical evidence to justify the analysis above by comparing the dynamics using (i) the effective two-level master equation derived in Eq.~\eqref{me1}, (ii) the standard bosonic master equation with $\dot{\hat{\rho}}=-i[\hat{H}_\mathrm{pcc},\hat{\rho}]+\kappa_\mathcal{C}\mathcal{D}[\hat{a}]\hat{\rho}$, and (iii) the master equation of a PCC coupled with a finite-linewidth cavity which emulates a general non-Markovian bath~\cite{imamog1994stochastic,xue2015quantum}. In case (iii) we assume that the PCC and the bath-cavity have the same frequency, so that the Hamiltonian of the system in the rotating-wave approximation (r.w.a) is, $\hat{H}_\mathrm{pcc,bc}=\hat{H}_\mathrm{pcc}+g(\hta^\dag\hta_\mathrm{bc}+\hta\hta^\dag_\mathrm{bc})$.
The linewidth of the bath-cavity is $\kappa_\mathrm{bc}$ and the system evolves according to the master equation, $\dot{\hat{\rho}}=-i[\hat{H}_\mathrm{pcc,bc},\hat{\rho}]+\kappa_\mathrm{bc}\mathcal{D}[\hta_\mathrm{bc}]\hat{\rho}$. To emulate the bath, we limit ourselves to the weak coupling regime $g\ll \kappa_\mathrm{bc},4K|\beta|^2$. In this limit, the master equation for the PCC, obtained by adiabatically eliminating the bath cavity, is of the form given in Eq.~\eqref{me1} with $\kappa'_\mathcal{C}=4g^2/\kappa_\mathrm{bc}$. Figure~\ref{dissipation_comp} shows numerical estimates for the probability of a bit-flip $\catmba\hat{\rho}_\mathrm{pcc}\catma$ and phase-flip $\bra{-}\hat{\rho}_\mathrm{pcc}\ket{-}$ error when the PCC is initialized in the cat state $\catpa$ or the superposition state $\ket{+}$ respectively (here $\ket{\pm}=(\catpa\pm\catma)/\sqrt{2}$ and $\hat{\rho}_\mathrm{pcc}$ is the reduced density matrix of the PCC). The magnitude of $\kappa_\mathcal{C}$ is same in (i) and (ii), while the parameters $g=0.05K$, $\kappa_\mathrm{bc}=2K$ in (iii) are chosen so that $\kappa'_\mathcal{C}=\kappa_\mathcal{C}$. 

The three different cases (i), (ii) and (iii) are depicted as solid lines, dots and triangles respectively. The value of $\beta$ is increased from $\beta=1$ in Fig.~\ref{dissipation_comp}(a) to $\beta=\sqrt{2}$ in Fig.~\ref{dissipation_comp}(b) and $\beta=2$ in Fig.~\ref{dissipation_comp}(c). As expected, since $4K\beta^2$ is large all the three cases give the same probability of bit- and phase-flip errors.  The probability of bit-flip error increases with $\beta$ whereas that of phase-flip error decreases with $\beta$. For example, the probability of phase-flip error at $t=2/\kappa_\mathcal{C}$ decreases from 0.018 in Fig.~\ref{dissipation_comp}(a) to 0.0067 and $5\times 10^{-7}$ in Fig.~\ref{dissipation_comp}(b) and Fig.~\ref{dissipation_comp}(c) respectively. Therefore the numerical simulations confirm that the probability of phase-flip error decreases exponentially with $\beta$.

 \subsection{Master equation with single-photon gain}
\begin{figure}
 \centering
 \includegraphics[width=\columnwidth]{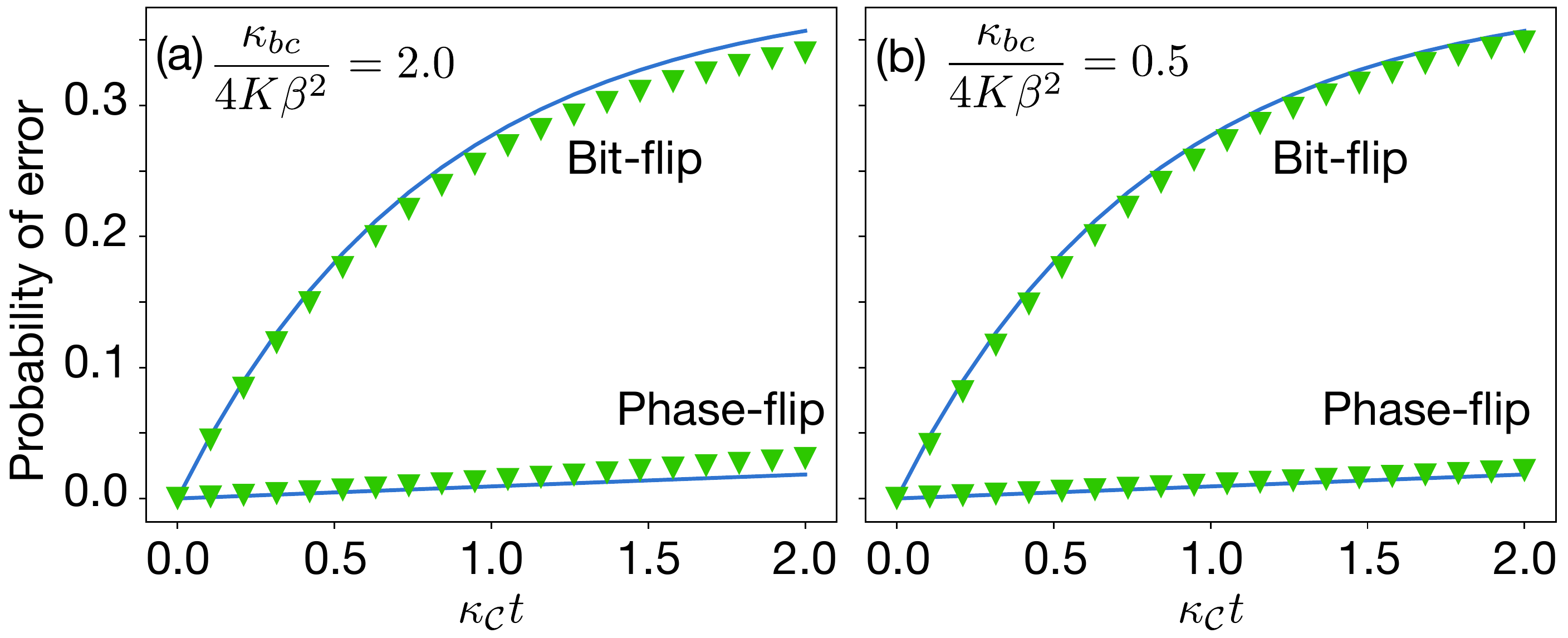}
 \caption{Comparison of dissipation predicted by (i) the effective two-level master equation in Eq.~\eqref{me_th} (solid line) and (ii) the master equation of a PCC coupled with a finite-linewidth cavity given by in Eq.~\eqref{me_bc} with $n_{th}=0.1$ (triangles). In both the cases $P=K$ so that $\beta=1$ and in (ii) $g=0.05K$. The linewidth $\kappa_{bc}$ of the cavity is $\kappa=0.5\times(4K\beta^2)$ in (a) and $\kappa=2.0\times(4K\beta^2)$ in (b). Note that $\kappa_\mathcal{C}(\omega_c)=4g^2/\kappa_{bc}$, so that $\kappa_\mathcal{C}(\omega_c)=0.005K$ in (a) and $\kappa_\mathcal{C}(\omega_c)=0.00125K$ in (b). The probability of bit-flip error is estimated by initializing the PCC to the state $\catpa$ and estimating $\catmba\hat{\rho}_\mathrm{pcc}\catma$. The probability of phase-flip error is estimated by initializing the PCC to the state $\ket{+}$ and estimating $\bra{-}\hat{\rho}_\mathrm{pcc}\ket{-}$. 
}
 \label{dissipation_thm}
 \end{figure}

 From the discussion in the previous section it is clear that while thermal photons at $\omega_\mathrm{pcc}$ can only cause transitions within $\mathcal{C}$, those at frequencies $\sim \omega_\mathrm{pcc}-\omega_\mathrm{gap}$ can cause excitations out of $\mathcal{C}$. In other words, the two-level approximation is strictly valid when the thermal noise-spectral density is colored or non-Markovian such that thermal photons at $\omega_\mathrm{pcc}-\omega_\mathrm{gap}$ are negligible. In this case, the effective two-level master equation becomes, 
\begin{align}
\dot{\hat{\rho}}&=-i[\hat{H}_\mathrm{pcc},\hat{\rho}]\nonumber\\
&+\kappa_\mathcal{C}(\omega_\mathrm{pcc})(1+n_\mathrm{th}(\omega_\mathrm{pcc}))\beta^2\mathcal{D}\left[p\catpa\catmba\right.\nonumber\\
&\left.+p^{-1}\catma\catpba\right]\hat{\rho}\nonumber\\
&+\kappa_\mathcal{C}(\omega_\mathrm{pcc})n_\mathrm{th}(\omega_\mathrm{pcc})\beta^2\mathcal{D}\left[p\catma\catpba+p^{-1}\catpa\catmba\right]\hat{\rho}.
\label{me_th}
\end{align}
The above equation was derived under the assumption that the spectral density of the environment is smooth or flat around $\omega_\mathrm{pcc}$, but falls off at $\omega_\mathrm{pcc}-\omega_\mathrm{gap}$. This analysis can be numerically confirmed by emulating such a bath with a finite-linewidth cavity which is coupled to the PCC. To ensure that the thermal photons in the bath do not excite the $\mathcal{C}_\perp$ subspace of the PCC, the linewidth of the bath cavity $\kappa_\mathrm{bc}$ is chosen to be smaller than $\omega_\mathrm{gap}$. The dynamics of such a system is described by the master equation, 
\begin{align}
\dot{\hat{\rho}}=-i[\hat{H}_\mathrm{pcc,bc},\hat{\rho}]+\kappa_\mathrm{bc}(1+n_\mathrm{bc})\mathcal{D}[\hta_\mathrm{bc}]\hat{\rho}+\kappa_\mathrm{bc}n_\mathrm{bc}\mathcal{D}[\hta_\mathrm{bc}]\hat{\rho}\label{me_bc},
 \end{align}
with $\hat{H}_\mathrm{pcc,bc}=\hat{H}_\mathrm{pcc}+g(\hta^\dag\hta_\mathrm{bc}+\hta\hta^\dag_\mathrm{bc})$. When $g\ll\kappa_\mathrm{bc}<\sim \omega_\mathrm{gap}$, the cavity emulates a non-Markovian bath. In other words, adiabatic elimination of the bath cavity would give Eq.~\eqref{me_th} with $\kappa'_\mathcal{C}(\omega_\mathrm{pcc})=4g^2/\kappa_{bc}$. Figure~\ref{dissipation_thm} compares the dynamics given by Eq.~\eqref{me_bc} and Eq.~\eqref{me_th} with $n_\mathrm{th}(\omega_\mathrm{pcc})=n_\mathrm{bc}$ and $\kappa_\mathcal{C}(\omega_\mathrm{pcc})=\kappa'_\mathcal{C}(\omega_\mathrm{pcc})=4g^2/\kappa_{bc}$. The numerical estimates for the probability of a bit-flip $\catmba\hat{\rho}_\mathrm{pcc}\catma$ and phase-flip $\bra{-}\hat{\rho}_\mathrm{pcc}\ket{-}$ error when the PCC is initialized in the cat state $\catpa$ or the superposition state $\ket{+}$ respectively are shown. The bath cavity is initialized to a thermal state with $n_\mathrm{bc}=0.1$. The coupling $g=0.05K$, $P=K$ ($\beta=1$), while $\kappa_\mathrm{bc}=2\times (4K\beta^2)$ in Fig.~\ref{dissipation_thm}(a) and $\kappa_{bc}=0.5\times (4K\beta^2)$ in Fig.~\ref{dissipation_thm}(b). As expected, the dynamics described by Eq.~\eqref{me_bc} and Eq.~\eqref{me_th} agree well (the solid lines and triangles overlap) for small $\kappa_\mathrm{bc}$ (Fig.~\ref{dissipation_thm}(b)) because the probability of excitations in the $\mathcal{C}$ subspace is small. However, as $\kappa_\mathrm{bc}$ increases (Fig.~\ref{dissipation_thm}(a)), the excitations in $\mathcal{C}_\perp$ become significant and the effective two-level ME in Eq.~\eqref{me_th} is no longer accurate. Note that the timescale for stabilizer measurement using the PCC is typically in the range of $T=1/K-10/K$. Therefore the relevant timescale for the plots in Fig.~\ref{dissipation_thm}(a,b) are $\kappa_\mathcal{C} T=0.001-0.05$. 

It is quite possible that spurious thermal excitations exist at the gap frequency or sudden non-perturbative effects cause excitations into the $\mathcal{C}_\perp$ subspace. These excitations, although rare, can impede the fault-tolerance of syndrome measurements. However as we will show in the next section, two-photon loss events will bring the system back to $\mathcal{C}$, thereby autonomously maintaining fault-tolerance.


\subsection{Autonomous correction of out-of-subspace excitation with two-photon dissipation}
This type of noise is present when photons are lost to the environment in pairs and will be invariably introduced when two-photon driving is applied to the PCC~\cite{wolinsky1988quantum,gilles1994generation,mirrahimi2014dynamically}. The rate of two-photon loss is typically negligible, but it can engineered to be larger~\cite{mirrahimi2014dynamically,leghtas2015confining,touzard2018coherent}. It arises when the system-bath coupling is of the form $
\hat{H}_\mathrm{pcc,b}=\sum_k g_{2,k}(\hta^2\htb^\dag_k e^{i(\omega_k-2\omega_\mathrm{pcc})t}+\hta^{\dag 2}\htb_ke^{-i(\omega_k-2\omega_\mathrm{pcc})t})$. The cat states are eigenstates of $\hta^2$ $\left(\hta^2\catpma=\beta^2\catmpa\right)$, but $\hta^{\dag 2}$ can excite the PCC to the $\mathcal{C}_\perp$ subspace. Following the discussion in the previous sections, it is evident that if the two-photon coupling with the environment is small, then excitations out of $\mathcal{C}$ will be negligible. Remarkably this implies that two-photon dissipation does not introduce any errors in the PCC. Moreover two-photon dissipation can bring spurious excitations in $\mathcal{C}_\perp$ back into $\mathcal{C}$. This can be understood through quantum-Zeno dynamics~\cite{facchi2002quantum,raimond2010phase,mirrahimi2014dynamically,cohen2017degeneracy,cohen2017autonomous} induced by the environment which constantly monitors the PCC with the two-photon process and projects it on to the $\mathcal{C}$ subspace. 

\begin{figure}
 \centering
 \includegraphics[width=.8\columnwidth]{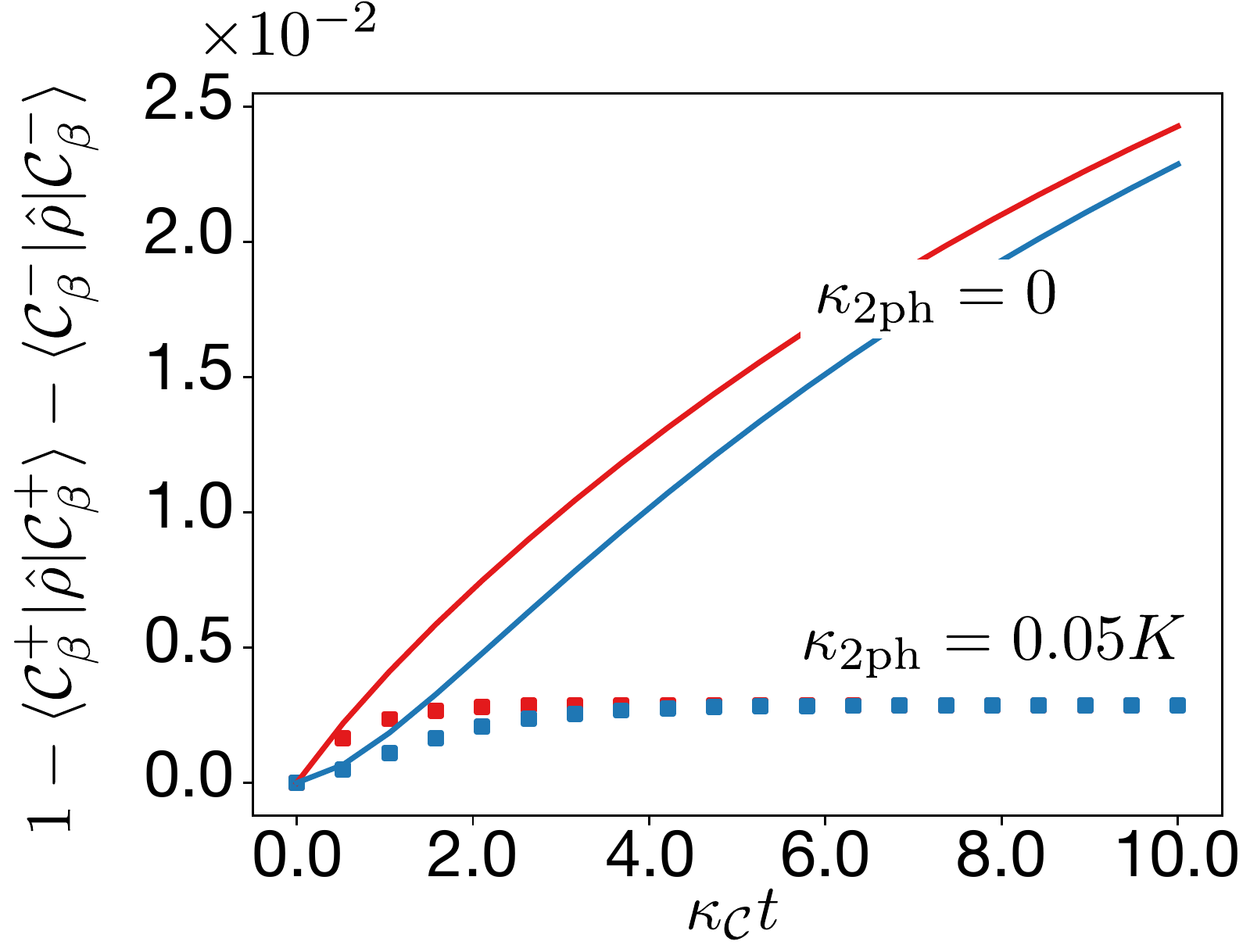}
 \caption{Probability of excitations out of $\mathcal{C}$ subspace, given by $1-\langle \mathcal{C}_\beta^+|\hat{\rho}|\mathcal{C}_\beta^+\rangle-\langle \mathcal{C}_\beta^-|\hat{\rho}|\mathcal{C}_\beta^-\rangle$, obtained by simulating Eq.~\eqref{me_bc} with $g=0.05K$, $P=K$ ($\beta=1$), $\kappa_\mathrm{bc}=8K$, $n_\mathrm{th}=0.1$ and an additional two-photon dissipation (rate $\kappa_\mathrm{2ph}$). The blue lines and blue dots correspond to when the PCC is initialized in the cat state $\catpa$ with $\kappa_\mathrm{2ph}=0$ and $\kappa_\mathrm{2ph}=0.05K$ respectively. The red lines and red dots correspond to when the PCC is initialized in the superposition state $\ket{+}$ with $\kappa_\mathrm{2ph}=0$ and $\kappa_\mathrm{2ph}=0.05K$ respectively. In practice, the stabilizer measurements typically would take time $T=1/K-10/K$ and so timescales such that $\kappa_\mathcal{C} T=0.005-0.05$ are relevant. 
}
 \label{thm_kapp2}
 \end{figure}

To elaborate with an example, suppose an excitation out of $\mathcal{C}$ is caused by a photon-gain event. The cat states transform as $\hta^\dag\catpma=\beta\catmpa+\ket{\psi^\pm_\perp}$ where $\ket{\psi^\pm_\perp}=\left(D(\beta)\ket{1}\pm D(-\beta)\ket{1}\right)/\sqrt{2}$ is a state in the $\mathcal{C}_\perp$ subspace. Here $D(\beta)$ is the displacement operator $D(\beta)=\exp\{\beta(\hta^\dag-\hta)\}$ and for convenience, we have approximated $\exp(-2\beta^2)\sim 0$. Therefore, the fraction of excitations in $\mathcal{C}_\perp$ $\propto1/\beta$ decreases as the size of the cat increases. Suppose a two-photon loss event occurred after a photon gain event, in which case the cat states transform as $\hta^2\hta^\dag\catpma=(\hta^\dag\hta^2+2\hta)\catpma=(\beta^3+2\beta)\catmpa+\beta^2\ket{\psi^\pm_\perp}$. In this case the fraction of excitations in $\mathcal{C}_\perp$ is $1/(\beta+2/\beta^2)<1/\beta$ and we find that the two-photon loss event has decreased the out of subspace excitations. The two-photon loss channel therefore is actually desirable because it autonomously corrects for out-of-subspace excitations in the PCC (of course, only as long as the rate of two-photon excitations is smaller than the gap between $\mathcal{C}$ and $\mathcal{C}_\perp$). 
To confirm this we simulate the dynamics in Eq.~\eqref{me_bc} with an additional two-photon dissipation $\kappa_\mathrm{2ph}\mathcal{D}[\hta^2]\hat{\rho}$, with $g=0.05K$, $P=K$ ($\beta=1$), $\kappa_\mathrm{bc}=8K$ and $n_\mathrm{th}=0.1$. Figure~\ref{thm_kapp2} shows the probability of excitations in $\mathcal{C}_\perp$ for $\kappa_\mathrm{2ph}=0$ (solid) and $\kappa_\mathrm{2ph}=0.05K$ (dotted), when the PCC is initialized in the cat state $\catpa$ (blue) or the superposition state $\ket{+}$ (red). As expected the out of subspace excitations decrease with increase in $\kappa_\mathrm{2ph}$. Because of finite $\kappa_\mathrm{2ph}/\omega_\mathrm{gap}$, the two-photon coupling with the environment can itself cause excitations out of $\mathcal{C}$. Therefore, the autonomous correction of out-of-subspace excitation with two-photon dissipation is not perfect (red and blue dots saturate to $\sim 3\times 10^{-4}$ in Fig.~\ref{thm_kapp2}). Note that a single-photon loss after a single-photon gain event will also decrease the excitations in $\mathcal{C}_\perp$, but it will also introduce bit-flips which, although can be overcome by majority vote, is less desirable.

\subsection{Master equation with pure dephasing}
In addition to the single-photon loss, gain and two-photon loss, it is possible that the frequency of the PCC fluctuates because of couplings with the environment which could be of the form,
$\hat{H}_\mathrm{pcc,b}=\sum g_{\phi,k}\hta^\dag\hta\htb^\dag_k\htb_k$. In this expression $\hat{b}_k$ are the bath modes. Following the discussions above we find that if $g_{\phi,k}\beta$ is smaller than $\omega_\mathrm{gap}$, then the master equation for this pure dephasing channel can be derived as
\begin{align}
\dot{\hat{\rho}}&=-i[\hat{H}_\mathrm{pcc},\hat{\rho}]\nonumber\\
&+\kappa_{\phi,\mathcal{C}}\beta^4\mathcal{D}\left[\left(\frac{p^2+p^{-2}}{2}\right)\hat{\tilde{\mathcal{I}}}-\left(\frac{p^{-2}-p^2}{2}\right)\hat{\tilde{\sigma}}_\mathrm{z}\right]\hat{\rho}.
\label{me_deph}
\end{align}

Note that $\hta^\dag\hta\ket{\pm\beta}=\beta^2\ket{\pm\beta}\pm\beta D(\pm\beta)\ket{1}$, so that the probability to go out of the $\mathcal{C}$ subspace is $(g_{\phi,k}\beta/\omega_\mathrm{gap})^2$. Therefore for this probability to be small, $g_{\phi,k}\beta\ll \omega_\mathrm{gap}$. Moreover, for $\beta\rightarrow 0$ the probability to go out of the cat subspace $\mathcal{C}$ vanishes. This is expected because $\beta=0$ corresponds to the case when the cat states reduce to the Fock states $\ket{n=0}$ and $\ket{n=1}$. In this case dephasing does not take the system out of $\mathcal{C}$, because Fock states are eigenstates of $\hta^\dag\hta$.

\begin{figure}
 \centering
 \includegraphics[width=.8\columnwidth]{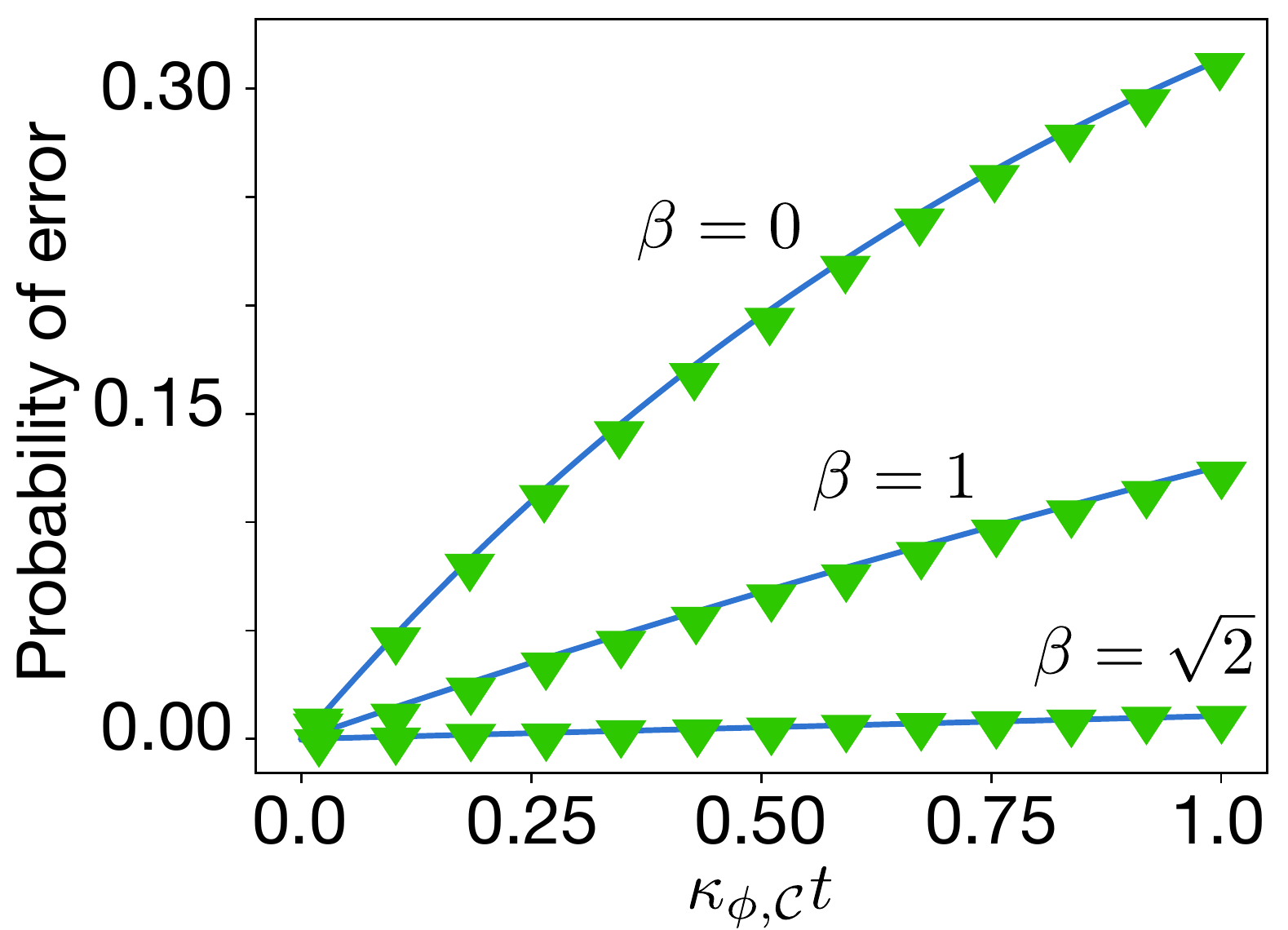}
 \caption{Comparison of dephasing predicted by (i) the effective two-level master equation in Eq.~\eqref{me_deph} (solid line) and (ii) the master equation of a PCC coupled with a finite-linewidth cavity given by in Eq.~\eqref{me_deph_bc} with $n_\mathrm{th}=1$ (triangles). In all the cases $K$ is fixed while $P$ is varied so that $\beta$ changes from $0$ to $1$ and $\sqrt{2}$. The parameters are chosen as $g=0.0025K$, $n_\mathrm{th}=1$, $\kappa_\mathrm{bc}=0.05K$ so that $\kappa'_{\phi,\mathcal{C}}=\kappa_{\phi,\mathcal{C}}=0.0005K$. The probability of phase-flip error is estimated by initializing the PCC to the state $\ket{+}$ and estimating $\bra{-}\hat{\rho}_\mathrm{pcc}\ket{-}$. For example, the probability of phase-flip errors for $\beta=\sqrt{2}$ at $t=1/\kappa_{\phi,\mathcal{C}}$ is $0.0125$. In practice, the stabilizer measurements typically take time $T=1/K-10/K$ and so the relevant timescales are $\kappa_{\phi,\mathcal{C}} T=0.0005-0.005\ll 1$ are relevant. }
 \label{deph_comp}
 \end{figure}

We justify the above master equation by comparing the dynamics using (i) the effective two-level master equation derived in Eq.~\eqref{me_deph} and (ii) the master equation of a PCC coupled with a cavity which emulates a general non-Markovian bath, with the Hamiltonian $\hat{H}_\mathrm{pcc,bc}=\hat{H}_\mathrm{pcc}+g\hta^\dag\hta(\hta^\dag_\mathrm{bc}\hta_\mathrm{bc}-\langle\hta^\dag_\mathrm{bc}\hta_\mathrm{bc}\rangle)$. Because of such an interaction, photon-number fluctuations in the bath cavity will cause fluctuations in the frequency of the PCC or in other words pure-dephasing. To emulate this effect we evolve the system according to the master equation, 
\begin{align}
\dot{\hat{\rho}}=-i[\hat{H}_\mathrm{pcc,bc},\hat{\rho}]+\kappa_\mathrm{bc}(1+n_\mathrm{th})\mathcal{D}[\hta_\mathrm{bc}]\hat{\rho}+\kappa_\mathrm{bc}n_\mathrm{th}\mathcal{D}[\hta^\dag_\mathrm{bc}]\hat{\rho}.
\label{me_deph_bc}
\end{align}
The fluctuations in the number of photons in the bath cavity becomes $n_\mathrm{th}+n_\mathrm{th}^2$. We limit the dynamics to the weak coupling limit $g\ll \kappa_\mathrm{bc}$ (so that the bath cavity indeed acts as a reservoir) and $g\ll4K|\beta|^2$ so that the two-level approximation is valid. In this limit, we expect the master equation for the PCC, obtained by adiabatically eliminating the bath cavity, to be of the form given in Eq.~\eqref{me_deph} with 
\be
\kappa'_{\phi,\mathcal{C}}=2g^2(n_\mathrm{th}+n_\mathrm{th}^2)/\kappa_\mathrm{bc}.
\ee
Figure~\ref{deph_comp} shows numerical estimates for the probability of a phase-flip $\bra{-}\hat{\rho}_\mathrm{pcc}\ket{-}$ error when the PCC is initialized in the superposition state $\ket{+}$ (here $\ket{\pm}=(\catpa\pm\catma)/\sqrt{2}$ and $\hat{\rho}_\mathrm{pcc}$ is the reduced density matrix of the PCC). In (ii) the parameters chosen are  $g=0.0025K$, $n_\mathrm{th}=1$, $\kappa_\mathrm{bc}=0.05K$ so that $\kappa'_\mathcal{C}=\kappa_\mathcal{C}=0.0005K$. The two different cases (i) and (ii) are depicted as solid lines and triangles respectively. The value of $\beta$ is increased from $\beta=0$ to $\beta={1}$ and $\beta=\sqrt{2}$. As expected, increase in $\beta$ exponentially suppresses the phase-flip rate.

\subsection{ Four-qubit stabilizer $\sxx{1}\sxx{2}\sxx{3}\sxx{4}$ in toric codes}
Extension of section IV.A makes it clear that it is also possible to measure the four-qubit stabilizer $\hat{S}_\mathrm{x}=\sxx{1}\sxx{2}\sxx{3}\sxx{4}$. The required interaction Hamiltonian between the qubits and PCC is $\hat{H}_\mathrm{I}=\chi(\sxx{1}+\sxx{2}+\sxx{3}+\sxx{4})(\hta^\dag+\hta)$. Such a coupling can be effectively implemented by the typical Jaynes-Cummings (JC) interaction given by $\hat{H}_\mathrm{I}=\chi\sum_i(\hta^\dag\smm{i}+\hta\spp{i})$. For $\chi$ smaller than the energy gap $\omega_\mathrm{gap}$, the PCC remains within $\mathcal{C}$ and the JC Hamiltonian reduces to $\hat{H}_\mathrm{I}=\chi\beta\sum_i(\sxx{i}(p+p^{-1})\tilde{\sigma}_\mathrm{x}/2+\sy{i}(p-p^{-1})\tilde{\sigma}_\mathrm{y}/2)$. For even moderately large amplitude of the cat state in PCC (such as $\beta=2$), the last term in the above equation $(\propto  p-p^{-1})$ becomes exponentially small and the desired interaction Hamiltonian for the measurement of $\hat{S}_\mathrm{x}$ stabilizer is obtained.

\begin{figure*}[ht]
 \centering
 \includegraphics[width=2\columnwidth]{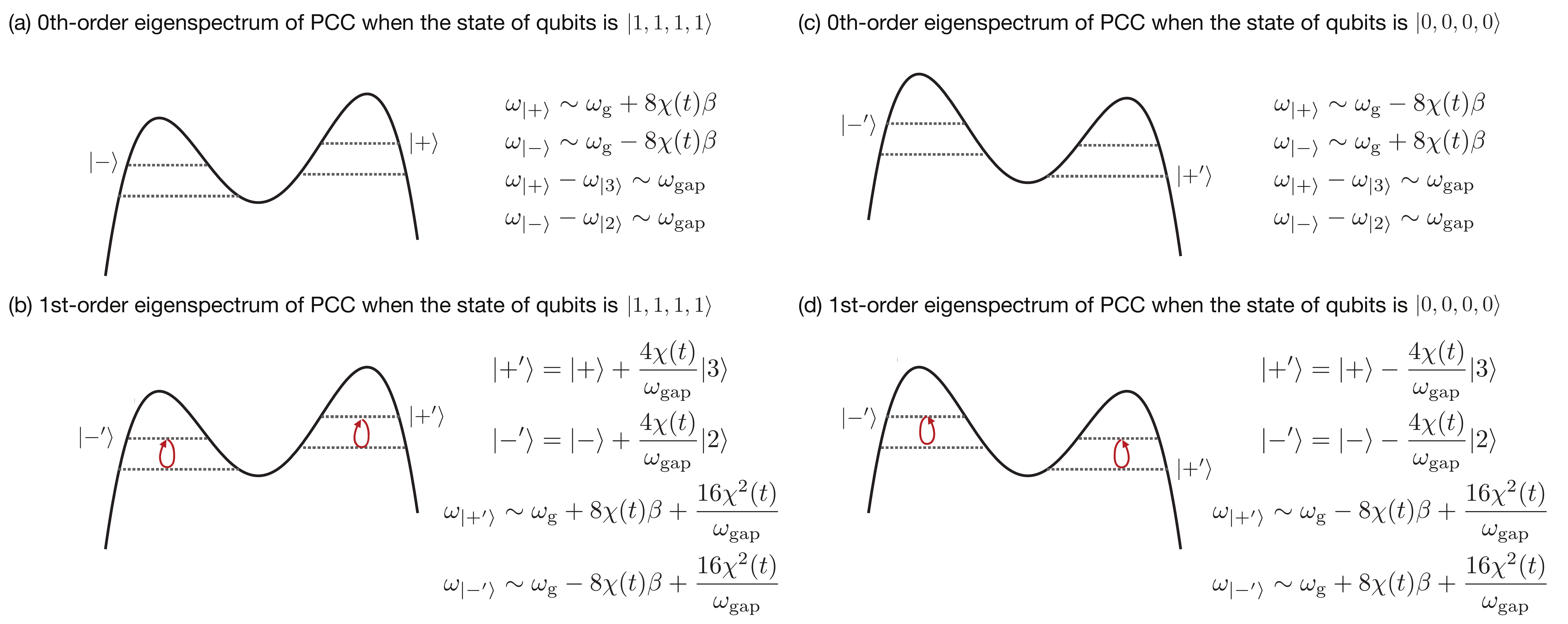}
 \caption{ The figure illustrates the eigenspectrum of the PCC while it is interacting with qubits. The solid line represents the inverted-double well structure of the meta-potential while the dotted lines represent the energy levels. Recall that, $\ket{2}\sim D(-\beta)\ket{n=1}$ and $\ket{3}\sim D(\beta)\ket{n=1}$. Unlike the non-interacting case (Fig.~\ref{cartoon1}), the qubit-PCC coupling lifts the degeneracy between $\ket{+}$ and $\ket{-}$. In other words, the meta-potential tilts to the left or right depending on whether the qubit state is (a) $\ket{1,1,1,1}$ or (c) $\ket{0,0,0,0}$. This is the leading-order effect of the interaction and the energies of the states $\ket{\pm}$ are $\omega_{\ket{\pm}}=\omega_\mathrm{g}\pm 8\chi(t)\beta$ in (a) and $\omega_{\ket{\pm}}=\omega_\mathrm{g}\mp 8\chi(t)\beta$ in (c). Importantly, if the energy gap is large, then the detuning between $\ket{+}$, $\ket{3}$ and that between $\ket{-}$, $\ket{2}$ remains the same ($\omega_{\ket{+}}-\omega_{\ket{3}}\simeq \omega_{\ket{-}}-\omega_{\ket{2}}\sim \omega_\mathrm{gap}$, in both (a,c)).  As shown in Appendix B, the effect of $\hta^\dag$ is to couple $\ket{+}$ to $\ket{3}$ and  $\ket{-}$ to $\ket{2}$. To 1$^\mathrm{st}$-order, in (b) $\ket{+'}=\ket{+}+(4\chi(t)/\omega_\mathrm{gap})\ket{3}$, $\ket{-'}=\ket{-}+(4\chi(t)/\omega_\mathrm{gap})\ket{2}$,  $\omega_{\ket{\pm'}}=\omega_\mathrm{g}\pm 8\chi(t)\beta+(16\chi(t)^2/\omega_\mathrm{gap})$. On the other hand, in (d) $\ket{+'}=\ket{+}-(4\chi(t)/\omega_\mathrm{gap})\ket{3}$, $\ket{-'}=\ket{-}-(4\chi(t)/\omega_\mathrm{gap})\ket{2}$, $\omega_{\ket{\pm'}}=\omega_\mathrm{g}\mp 8\chi(t)\beta+(16\chi(t)^2/\omega_\mathrm{gap})$. 
}
 \label{cartoon2}
 \end{figure*}

\subsection{Phase diffusion during the measurement of the $\hat{S}_\mathrm{z}$ stabilizer}
As discussed in the main text the qubit-cavity coupling $\propto\chi(t)(\hta^\dag+\hta)$ can cause a small virtual excitation of the states in $\mathcal{C}_\perp$. As $\chi(t)$ is turned off, these virtual excitations quickly reduce and the cavity returns to the cat manifold carrying with itself an extra phase. This extra phase, which depends on $\chi(t)$, is different for different qubit states and therefore leads to some cavity-qubit entanglement and hence phase diffusion. 

To elaborate, consider the eigenspectrum of the PCC when the state of the qubits is $\ket{1,1,1,1}$ shown in Fig.~\ref{cartoon2}. The results in this section are best understood by working in the $\ket{\pm}$ eigenbasis of the PCC where $\ket{\pm}=(\catpa\pm\catma)/\sqrt{2}$.
The coupling between the qubits and PCC is $\chi(t)(\sz{1}+\sz{2}+\sz{3}+\sz{4})(\hta^\dag+\hta-2\beta)$. As a result, when the qubits state is $\ket{1,1,1,1}$, the PCC experiences a single-photon drive of strength $4\chi(t)(\hta^\dag+\hta)$. This drive tilts the meta-potential of the PCC as shown in Fig.~\ref{cartoon2}(a). We will refer to this as the 0$^\mathrm{th}$-order effect. When $\chi(t)$ is small w.r.t to the energy gap $\omega_\mathrm{gap}$, the single-photon drive lifts the degeneracy between the $\ket{+}$ and $\ket{-}$ states by $16\chi(t)\beta$. This lifting of degeneracy arises because the single-photon drive couples the cat states $\catpa$ and $\catma$, which is what is exploited to perform the stabilizer measurements. To 0$^\mathrm{th}$ order, the enegies of the states $\ket{\pm}$ are $\omega_{\ket{\pm}}=\omega_\mathrm{g}\pm 8\chi(t)\beta$, where $\omega_\mathrm{g}$ is their energy when $\chi=0$. Importantly, when $\omega_\mathrm{gap}$ is large, the energy difference between $\ket{+}$, $\ket{3}$ and that between $\ket{-}$, $\ket{2}$ remains the same. That is, $\omega_{\ket{+}}-\omega_{\ket{3}}\simeq \omega_{\ket{-}}-\omega_{\ket{2}}\sim \omega_\mathrm{gap}$ or in other words, the tilting of the the meta-potential is uniform. Note that, this approximation is only valid when mixing with other states is negligible and breaks down with increase in $\chi$. \\
\indent
Recall the results of Appendix B in which we showed that $\hta^\dag$ couples the states $\ket{+/-}$ and $\ket{3/2}$. Therefore, the 1$^\mathrm{st}$-order effect of the single-photon drive, illustrated in Fig.~\ref{cartoon2}(b) is to mix these states, resulting in the time-dependent dressed states $\ket{+'}=\ket{+}+(4\chi(t)/\omega_\mathrm{gap})\ket{3}$ and $\ket{-'}=\ket{-}+(4\chi(t)/\omega_\mathrm{gap})\ket{2}$, with energies $\omega_{\ket{\pm'}}=\omega_\mathrm{g}\pm 8\chi(t)\beta+(16\chi(t)^2/\omega_\mathrm{gap})$. If the coupling strength $\chi(t)$ is tuned adiabatically then the state of the PCC follows the instantaneous eigenstates $\ket{\pm'}$. As a result, an initial state $\ket{1,1,1,1}\otimes \catpa$ evolves after a time $T_\mathrm{z}=\pi/(8\chi\beta)$ to $\exp(i\phi_{1,1,1,1}) \ket{1,1,1,1}\otimes \catpa$, where
\be
\phi_{1,1,1,1}=\int_0^{T_\mathrm{z}}\frac{16\chi(t)^2}{\omega_\mathrm{gap}} dt
\label{phase1}
\ee
In other words, after the duration of the stabilizer measurement protocol, $T_\mathrm{z}$, the state acquires an additional phase $\phi_{1,1,1,1}$. 
Following a similar argument (illustrated by the eigenspectrum analysis in Fig.~\ref{cartoon2}(c,d)) we find that, $\ket{0,0,0,0}\otimes \catpa$ evolves after a time $T_\mathrm{z}=\pi/(8\chi\beta)$ to $\exp(i\phi_{0,0,0,0}) \ket{0,0,0,0}\otimes \catpa$, where
\be
\phi_{0,0,0,0}=\int_0^{T_\mathrm{z}}\frac{16\chi(t)^2}{\omega_\mathrm{gap}} dt
\label{phase2}
\ee
Note that the phase is proportional to the square of the coupling in the 1$^\mathrm{st}$-order approximation and hence $\phi_{0,0,0,0}=\phi_{1,1,1,1}$. Since the coupling between the rest of the even parity states $\ket{0,0,1,1}$, $\ket{1,1,0,0}$, etc is zero $\phi_{0,0,1,1}=\phi_{1,1,0,0}=....=0$. This difference in the phases corresponding to the states $\ket{0,0,0,0}$, $\ket{1,1,1,1}$ and other even parity states $\ket{0,0,1,1}$, $\ket{1,1,0,0}$, etc., leads to phase diffusion when the qubits are initialized in a superposition state such as $\ket{\psi_\mathrm{e}}$ in Eq.~\eqref{even_q1}. That is, the overlap between the state of the qubits after time $T_\mathrm{z}$ and $\ket{\psi_\mathrm{e}}$ is not one. However, as long as the coupling is tuned adiabatically and for small $\chi/\omega_\mathrm{gap}$, this phase diffusion is small. Although to leading order $\phi_{1,1,1,1}=\phi_{0,0,0,0}$, this will not be the case as $\chi$ increases because of coupling with other states in the Hilbert space of the PCC. Similarly, according to our 1$^\mathrm{st}$-order theoretical analysis, there should be no phase diffusion when the qubits are initialized in the odd parity subspace. This is because the phase is proportional to the square of the coupling, which is the same for all the odd parity states $\ket{1,0,0,0}$, $\ket{1,1,1,0}$ etc (under the assumption that the $\chi$'s are equal). However, as $\chi$ increases, higher-order effects from other states in the PCC will have to be taken into account which will also lead to phase diffusion in the odd-parity subspace. 

\begin{figure}
 \centering
 \includegraphics[width=\columnwidth]{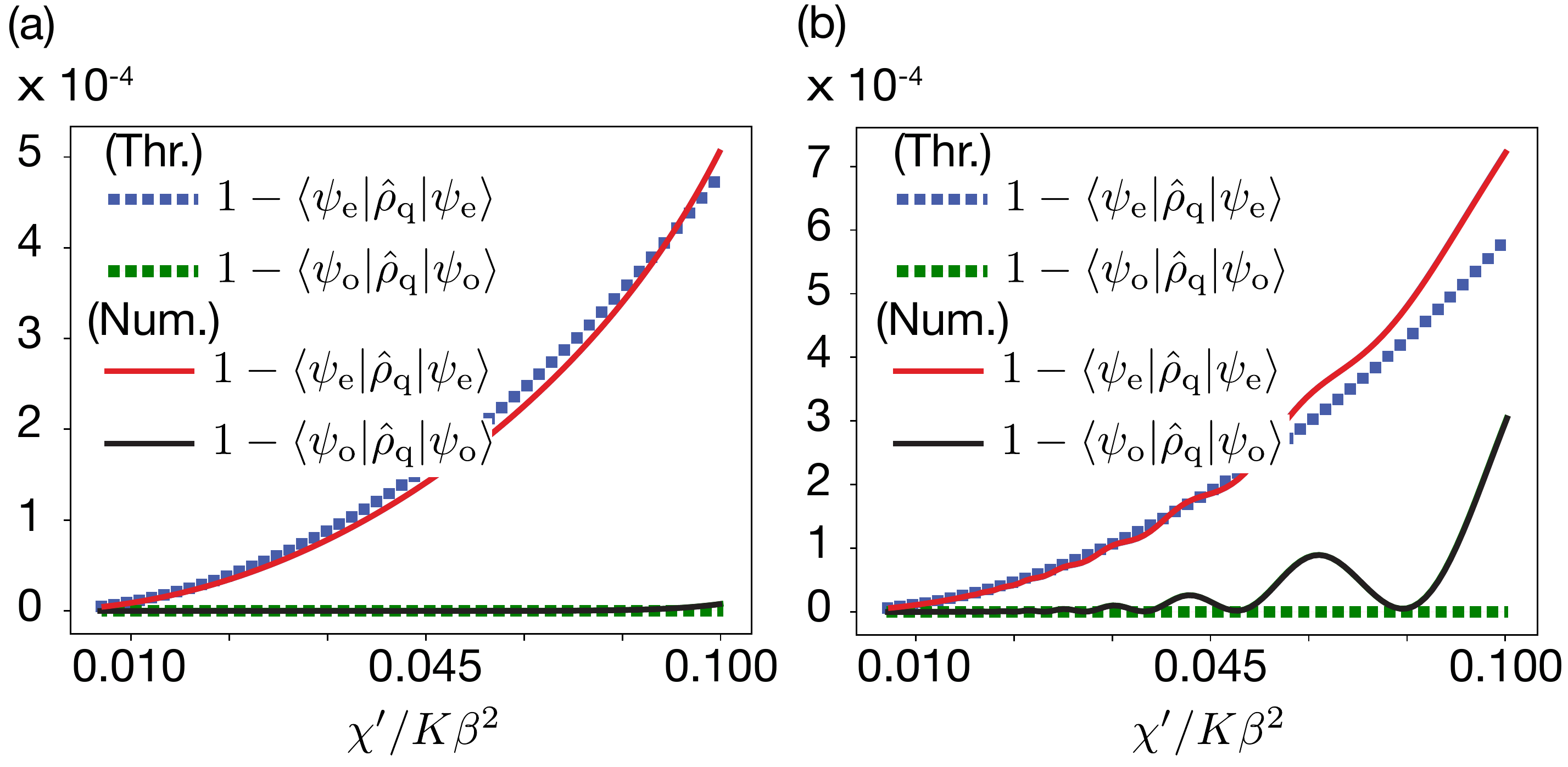}
 \caption{Dependence of phase diffusion in the state of the qubits as a function of the coupling strength. In (a) the time dependence of the interaction between the qubits and PCC is taken as Gaussian $\chi(t)=(\chi_0/\sqrt{\pi})\exp(-t^2/T_\mathrm{z}^2)$ with cut-offs at $\pm 3T_\mathrm{z}$.
In (b) the time dependence of the interaction between the qubits and PCC is $\chi(t)=0.5\chi_0\pi\sin(\pi t/T_\mathrm{z})$ with cut-offs at $0$ and $T_\mathrm{z}$. $\chi'$ is the peak interaction strength so that in (a) $\chi'=\chi_0\sqrt{\pi}$ and in (b) $\chi'=\pi\chi_0/2$. The parameters are $P=4K$ ($\beta=2$) and $T_\mathrm{z}=\pi/8\chi_0\beta$. The phase diffusion is suppressed when the excitations out of the $\mathcal{C}$ subspace are negligible, that is, when $\chi/'K\beta^2$ is small. Both theoretical (dotted lines) and numerical (solid lines) are shown. The agreement between the two is good for small $\chi'$. As $\chi$ increases the 1$^\mathrm{st}$-order perturbation theory is not sufficient and effect of other states in the PCC must be included. Also in (b), effect of non-adiabatic terms and higher-order effects become apparent at smaller $\chi'$ compared to (a). For small $\chi'$ the average phase diffusion decreases quadratically with $\chi'/K\beta^2$. For example, the phase diffusion is $<10^{-4}$ when $\chi'/K\beta^2<0.045$. }
 \label{toric1}
 \end{figure}

We now compare our simple theoretical prediction with numerical results. The qubits are initialized in the even parity state,
\begin{align}
\ket{\psi_\mathrm{e}}=\frac{1}{\sqrt{8}}\left(\sum_{i,j} \hat{\sigma}_{\mathrm{x},i}\hat{\sigma}_{\mathrm{x},j}+\hat{I}+\hat{\sigma}_{\mathrm{x},1}\hat{\sigma}_{\mathrm{x},2}\hat{\sigma}_{\mathrm{x},3}\hat{\sigma}_{\mathrm{x},4}\right)\ket{0,0,0,0}.
\end{align}
and the PCC is initialized in the cat state $\catpa$. The evolution under the Hamiltonian $H=-K\hta^{\dag 2}\hta^2+P(\hta^{\dag 2}+\hta^2)+\chi(t) \hat{S}'_\mathrm{z}(\hta+\hta^\dag-2\beta)$ is numerically simulated with $P=2K$, $\chi(t)=(\chi_0/\sqrt{\pi})\exp(-t^2/T_\mathrm{z}^2)$, and $T_\mathrm{z}=\pi/8\chi_0\beta$. The cut-offs for the Gaussian pulse are taken at $\pm3T_\mathrm{z}$. Figure~\ref{toric1}(a) shows the probability of phase-diffusion given by $E_\mathrm{e}=1-\langle\psi_\mathrm{e}|\hat{\rho}_\mathrm{q}|\psi_\mathrm{e}\rangle$ (solid red line) as a function of $\chi/K\beta^2$ (here $\chi$ is the peak interaction strength $\chi=\chi_0/\sqrt{\pi}$). The blue dotted line shows the theoretically estimated $E_\mathrm{e}$, using the formulas for phases derived in Eq.~\eqref{phase1} and Eq.~\eqref{phase2}. The phase diffusion is proportional to the square of the coupling, that is, $E_\mathrm{e}\propto \chi^2$ for small $\chi$.\\
\indent 
The analysis is repeated with the qubits initialized in the odd parity state,
\begin{align}
\ket{\psi_\mathrm{o}}=\frac{1}{\sqrt{8}}\left(\sum_i \hat{\sigma}_{\mathrm{x},i}+\sum_{i,j,k} \hat{\sigma}_{\mathrm{x},i}\hat{\sigma}_{\mathrm{x},j}\hat{\sigma}_{\mathrm{x},k}\right)\ket{0,0,0,0}.
\end{align}
Figure.~\ref{toric1}(a) shows the probability of phase-diffusion given by $E_\mathrm{o}=1-\langle\psi_\mathrm{o}|\hat{\rho}_\mathrm{q}|\psi_\mathrm{o}\rangle$ (solid black line). The 1$^\mathrm{st}$-order perturbation theory predicts an absence of phase-diffusion in this state. The green dotted line shows the theoretically estimated $E_\mathrm{e}=0$ as a reference. We find that the simple 1$^\mathrm{st}$-order theory agrees well with the numerical results for small $\chi$. However as $\chi$ increases, the disagreement between numerical estimates and theory increases, which is expected because the small $\chi$ approximation breaks down. \\
\indent
The Gaussian pulse shape chosen for the above example behaves well and ensures the adiabacity condition that we have assumed in the theory. However, we can test our theory against another pulse shapes, such as a sine pulse $\chi(t)=0.5\chi_0\pi\sin(\pi t/T_\mathrm{z})$ with cut-offs at $t=0$ and $t=T_\mathrm{z}$. In this case we expect non-adiabatic effects to emerge. We repeat the above analysis for this pulse with the results shown in Figure.~\ref{toric1}(b). The solid lines are from numerical simulations, while the dotted ones are from the 1$^\mathrm{st}$-order theory assuming adiabaticity. In this case, again we see agreement between theory and numerical results for small $\chi$. But the theoretical and numerical results deviate for larger $\chi$. The numerically obtained solid lines show some oscillations because of non-adiabatic effects. Moreover, deviations between the the solid and dashed lines become more prominent for the sine-pulse compared to that with a Gaussian pulse. \\
\indent
To summarize, we expect phase diffusion in the qubits states because of interaction with other states in the Hilbert space of the PCC. This phase diffusion is proportional to $(\chi/\omega_\mathrm{gap})^2$ (for small $\chi$) and can be suppressed by increasing the energy gap $\omega_\mathrm{gap}$.

\subsection{Phase diffusion during the measurement of $e^{i\pi\hta^\dag_\mathrm{s}\hta_\mathrm{s}}$ stabilizer}
\begin{figure}
 \centering
 \includegraphics[width=\columnwidth]{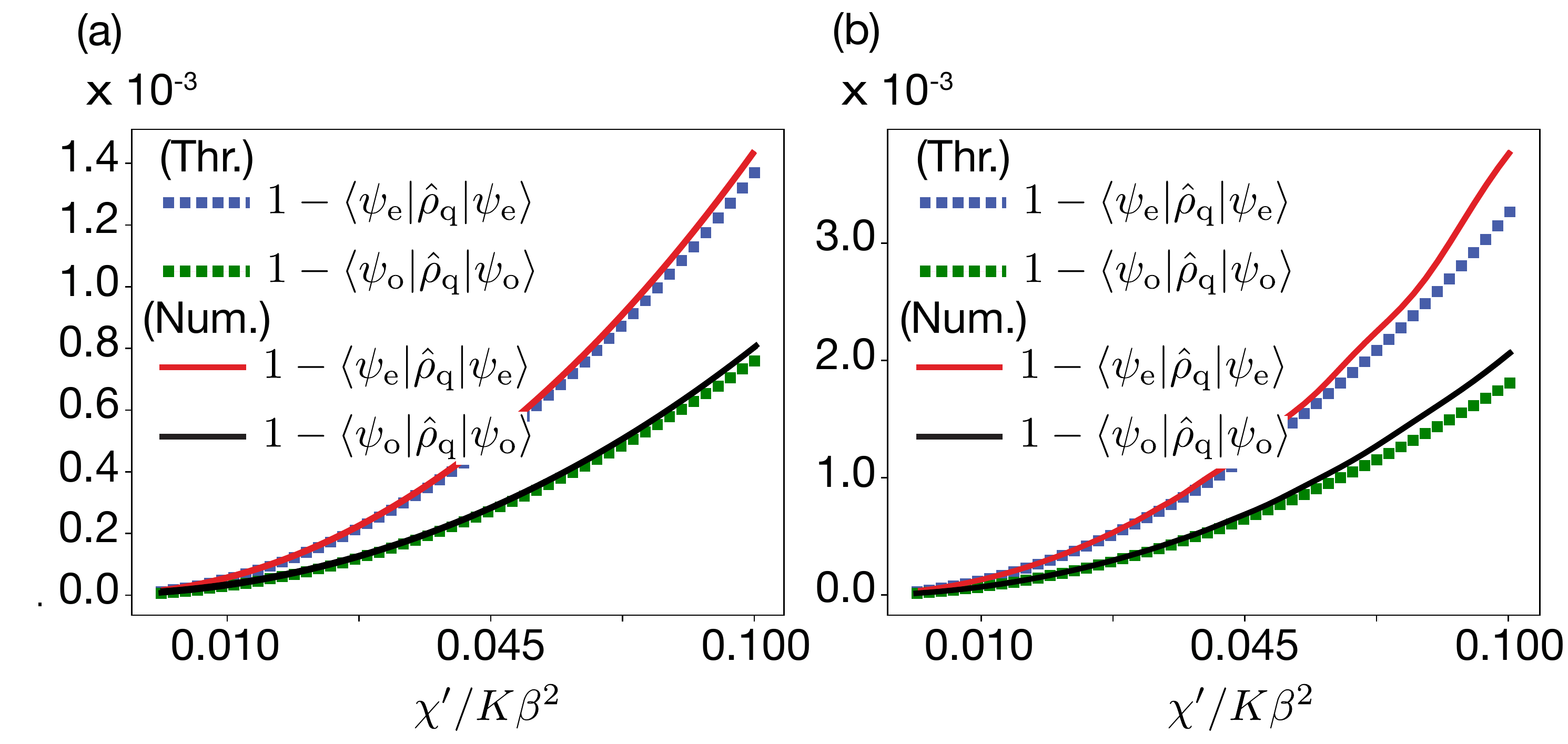}
  \caption{Dependence of phase diffusion in the state of the storage cat as a function of the coupling strength. In (a) the time dependence of the interaction between the storage and PCC is $\chi(t)=(\chi_0/\sqrt{\pi})\exp(-t^2/T_\mathrm{p}^2)$ with cut-offs at $\pm 3T_\mathrm{p}$.
In (b) the time dependence of the interaction is $\chi(t)=0.5\chi_0\pi\sin(\pi t/T_\mathrm{p})$ with cut-offs at $0$ and $T_\mathrm{p}$. $\chi'$ is the peak interaction strength so that in (a) $\chi'=\chi_0\sqrt{\pi}$ and in (b) $\chi'=\pi\chi_0/2$. The parameters are $P=4K$ ($\beta=2$) and $T_\mathrm{p}=\pi/4\chi_0\beta$. The phase diffusion is suppressed when the excitations out of the $\mathcal{C}$ subspace are negligible, that is, when $\chi/'K\beta^2$ is small. Both theoretical (dotted lines) and numerical (solid lines) are shown. The agreement between the two is good for small $\chi'$ and when the adiabatic approximation is valid. Again we find that for small $\chi'$ the average phase diffusion decreases quadratically with $\chi'/K\beta^2$. For example in (a), the phase diffusion is $<10^{-4}$ when $\chi'/K\beta^2<0.02$. 
 }
 \label{cat1}
 \end{figure}

Following the discussion in the previous section it is easy to see that the coupling between the storage-cavity and the PCC causes a small virtual excitation out of the cat subspace and leads to dressing of the $\ket{\pm}$ states.
The Fock states comprising the storage cat $\ket{n=0},\ket{n=1},\ket{n=2}...$ couple to the PCC with different strengths ($\propto n\chi$). If the coupling between the storage-cat and PCC is tuned adiabatically, then after a time $T_\mathrm{p}$ an initial state $\ket{2n}\otimes\catpa$ evolves to $\exp(i\phi_{2n})\ket{2n}\otimes\catpa$ and the state $\ket{2n+1}\otimes\catpa$ evolves to $\exp(i\phi_{2n+1})\ket{2n+1}\otimes\catma$. Here the phase dependent on storage-photon-number $m$ is,
\be
\phi_m=\int_0^{T_\mathrm{p}}\frac{m^2\chi(t)^2}{\omega_\mathrm{gap}} dt.
\label{phasem}
\ee
Recall that $T_\mathrm{p}$ is the time required to map the stabilizer on to the PCC.
This storage-photon-number-dependent phase leads to phase diffusion when the storage is in a superposition of Fock states such as a cat state. Full correction of phase diffusion would require complete knowledge of the photon statistics of the storage which would defeat the purpose of error correction. 
However, correction of the mean phase is possible by applying a counter-drive to the PCC as shown in Eq.~\eqref{catH}.  \\
\indent
We again compare the simple theoretical prediction with numerical results. The evolution under the following Hamiltonian is numerically simulated,
\begin{align}
\hat{H}&=-K\hta^{\dag 2}\hta^2+P(\hta^{\dag 2}+\hta^2)\nonumber\\
&+\chi(t) (\hta^\dag_\mathrm{s}\hta_\mathrm{s}-\langle\hta^\dag_\mathrm{s}\hta_\mathrm{s}\rangle)(\hta+\hta^\dag-\langle\hta+\hta^\dag\rangle),
\end{align}
and the reduced density matrix of the storage $\hat{\rho}_\mathrm{s}$ is obtained. Figure~\ref{cat1} shows the probability of phase diffusion when the storage is initialized in the even and odd parity states, $E_\mathrm{o}=1-\langle\psi_\mathrm{o}|\hat{\rho}_\mathrm{s}|\psi_\mathrm{o}\rangle$ and $E_\mathrm{e}=1-\langle\psi_\mathrm{e}|\hat{\rho}_\mathrm{s}|\psi_\mathrm{e}\rangle$. Here $\ket{\psi_\mathrm{o}}=\catm+i\catmis$ and $\ket{\psi_\mathrm{e}}=\catp+\catpis$. Both theoretical (dotted lines) and numerical (solid) lines are shown. The theoretical results are based on phases estimated in Eq.~\eqref{phasem}. Again we find good agreement between theory and numerical simulations for small $\chi$ for both the Gaussian pulse $\chi(t)=(\chi_0/\sqrt{\pi})\exp(-t^2/T_\mathrm{p}^2)$ (with cut-offs at $\pm 3T_\mathrm{p}$, Fig.~\ref{cat1}(a)) and the sine pulse $\chi(t)=0.5\chi_0\pi\sin(\pi t/T_\mathrm{p})$ (with cut-offs at $0$ and $T_\mathrm{p}$, Fig.~\ref{cat1}(b)). Here $T_\mathrm{p}=\pi/4\chi_0\beta$. Again we find that the average phase diffusion is suppressed for small $\chi/K\beta^2$ and the PCC indeed measures the stabilizer $\langle\hat{P}\rangle$ without revealing any information about the underlying photon number statistics.

\subsection{Estimating the feedback phases for phase estimation} 

To understand how the feedback phases $\phi$ and $\varphi$ are determined for the phase estimation protocol described in section IV.C, suppose that the storage was in the eigenstate of the stabilizer $\hat{S}_\mathrm{q}$ with eigenvalue $\exp(2i\sqrt{\pi}u)$. In this case, the state of the PCC after the application of the gate $\hat{U}_1(T_1)$ is $i\catma \sin(\sqrt{\pi}u)+\catpa \cos(\sqrt{\pi}u)$. If the PCC is further rotated around the $X$ axis of the Bloch sphere by an angle $\phi/2$ (by application of single-photon drive), its state becomes $i\catma \sin(\sqrt{\pi}u+\phi)+\catpa \cos(\sqrt{\pi}u+\phi)$. The probability for the PCC to remain in the $\catpa$ state after a round of phase estimation is $P_\phi(+|u)=\cos^2(\sqrt{\pi}u+\phi/2)$. Therefore, in order to accurately predict $u$, the sensitivity of the probability distribution
 $\partial P_\phi(+|u)/\partial\phi$ must be maximized. This is achieved in APE by choosing the feedback phase $\phi$ dependent on whether the PCC evolved to $\catpa$ or $\catma$ in the previous round of phase estimation. A similar analysis applies for the APE of the eigenvalue of $\hat{S}_\mathrm{p}$. In the simulations presented in the main text, the initial GKP state is approximately the eigenstate of the stabilizers with eigenvalues $u,v=0$. Furthermore, only one round of phase estimation is carried out. Therefore to maximize $\partial P_\phi(+|u)/\partial\phi$ we choose $\phi=\pi/2$. 
 
\subsection{Holevo phase variance in the presence of single-photon loss in the PCC} In this case the reduced density matrix of the storage $\hat{\rho}^\mathrm{m}_\mathrm{s}$ is obtained at time $T=\sqrt{\pi}/(g\beta\sqrt{2})$ by numerically solving the master equation in Eq.~\eqref{me_gkp} (i.e., without performing any rotations and measurements of the PCC). The variance is evaluated as ${V}^{\mathrm{m}}_\mathrm{p,q}=s^{-2}_\mathrm{p,q}-1$, with $s_\mathrm{p,q}=\mathrm{Tr}[\hat{S}_\mathrm{p,q}\hat{\rho}^\mathrm{m}_\mathrm{s}]$. Note that the variance evaluated this way is equivalent to throwing away the measurement result. Since the measurement results are discarded, no information about the storage cavity is obtained and the Holevo phase variance remains the same as that of the initial state $V^0_\mathrm{q,p}$.

\subsection{One round of APE with a qubit} One round of phase estimation for $\hat{S}_\mathrm{p}$ with an ideal two-level system is simulated using the master equation,
\be
\dot{\hat{\rho}}=-i[\hat{H},\hat{\rho}]+\gamma\mathcal{D}[\hat{\sigma}_-]\hat{\rho},\quad \hat{H}=ig_\mathrm{q}(\hta^\dag_\mathrm{s}-\hta_\mathrm{s})\hat{\sigma}_\mathrm{x}.
\label{me_gkp_qubit}
\ee 
Firstly, simulations are performed with $\gamma=0$. The density matrix for the system is obtained at time $T_\mathrm{ideal}=\sqrt{\pi}/(g_\mathrm{q}\beta)$. After this, the qubit is rotated around the $X$-axis by $\phi=\pi/2$, which is followed by projective measurement of the qubit along the $Z$-axis. The reduced density matrix for the storage cavity, $\hat{\rho}_{s,\pm }$ is obtained, from which the Holevo variances ${V'}^\mathrm{ideal}_\mathrm{q,p}$ are evaluated. Next simulations are performed with $\gamma\neq 0$. In this case, the reduced density matrix of the storage is obtained at time $T_\mathrm{ideal}$ (i.e., without performing any measurements on the ideal qubit). In this case, the variance is evaluated as ${V}^{\mathrm{m,ideal}}_\mathrm{p,q}=s^{-2}_\mathrm{p,q}-1$, with $s_\mathrm{p,q}=\mathrm{Tr}[\hat{S}_\mathrm{p,q}\hat{\rho}_\mathrm{s}]$.

\begin{figure}
 \centering
 \includegraphics[width=\columnwidth]{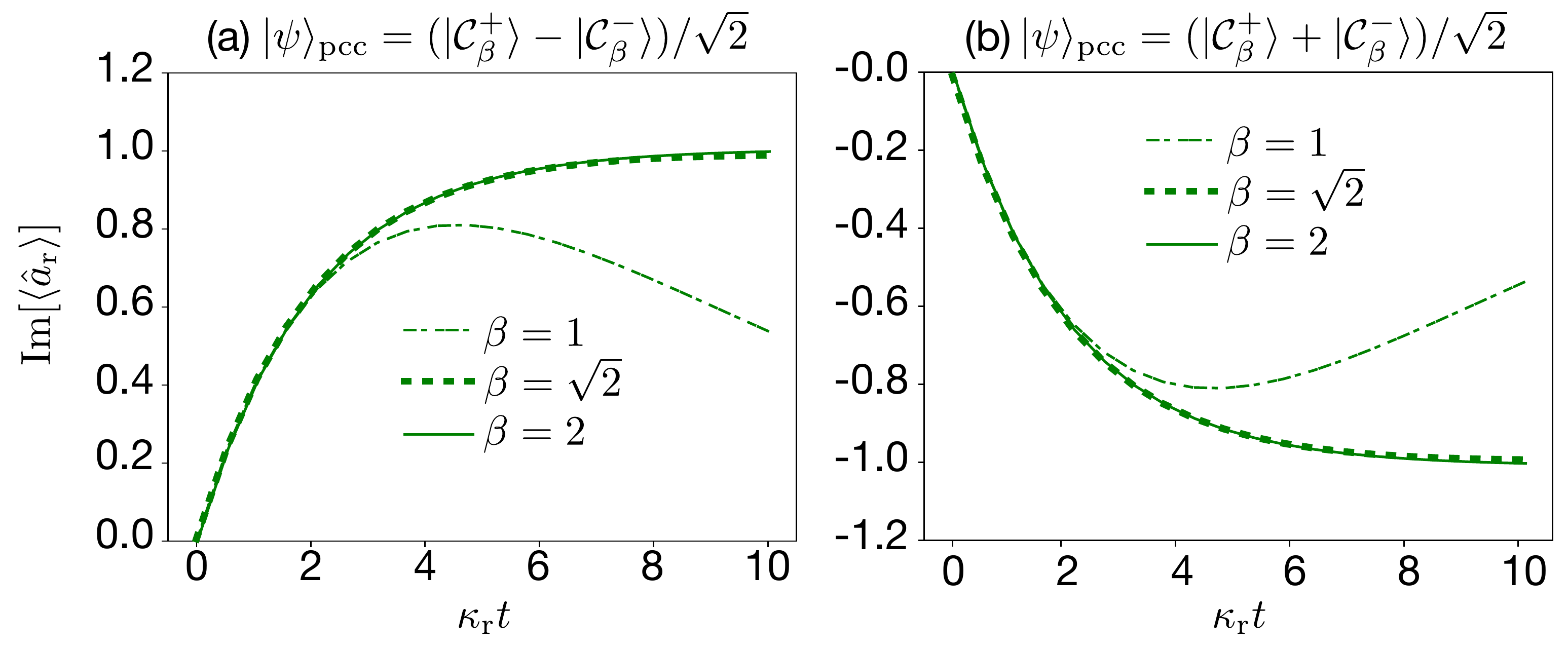}
 \caption{ Evolution of the $I-$quadrature ($i(\hta^\dag_\mathrm{r}-\hta_\mathrm{r})/2$) of the field in the readout cavity when the state of the PCC is (a) $(\catpa-\catma)/\sqrt{2}$ and (b) $(\catpa+\catma)/\sqrt{2}$, for $\beta=1,\sqrt{2},2$. Clearly the displacement of the readout resonator field is conditioned on the state of the PCC. }
 \label{read_fig}
 \end{figure}

\subsection{Errors during rotation under Kerr-evolution}
During free evolution under the Kerr term, single-photon loss can cause three effects - (a) Shrinkage of the size of the coherent states (this is not a debilitating effect because the states can be re-inflated simply by re-applying the pump), (b) bit-flips, and (c) additional phase rotations (due to non-commutativity of the Kerr-rotation and single-photon loss). Suppose a photon jump event happened at time $t_j$, then the state of the system is 
\begin{align}
&e^{iK(T-t_j)\hta^{\dag 2}\hta^2+iK(T-t_j)\hta^\dag\hta}\hta e^{iKt_j\hta^{\dag 2}\hta^2+iKt_j\hta^\dag\hta} (\catpa\pm i\catma)\nonumber\\
&=e^{2t_jK\hta^\dag\hta}e^{iKT\hta^{\dag 2}\hta^2+iKT\hta^\dag\hta}\hta (\catpa\pm i\catma)\nonumber\\
&=e^{2t_jK\hta^\dag\hta}e^{iKT\hta^{\dag 2}\hta^2}(\catpa\mp i\catma)\nonumber\\
&=e^{2t_jK\hta^\dag\hta}(\catpa\mp \catma)\nonumber\\
&={\ket{\mathcal{C}^+_{\beta e^{i\theta}}}}\mp {\ket{\mathcal{C}^-_{\beta e^{i\theta}}}}, \quad \theta=2t_jK.
\end{align}
Unless $\theta=n\pi$, the resulting states lie out side the $\mathcal{C}$ subspace. If the single photon loss takes place in the beginning of the protocol, (i.e., $\theta= 0$) then the readout merely gives an incorrect measurement result. This can be recovered while repeating the entire protocol a few times and taking a majority vote over the outcomes. If $\theta= \pi$, i.e., the photon loss takes place at the end of the protocol, then we recover the correct state. This is trivial because at the end of the protocol the two states are aligned along the $X$ axis and are invariant under the single-photon loss channel. On the other hand, if the photon loss happens at $t_j\neq 0, \pi/2K$ then we end up in a state outside the $\mathcal{C}$~\cite{mirrahimi2014dynamically}. The next step of Q-switching and homodyne measurement will reveal if this error happened, in the event of which, the PCC can be re-initialized. Moreover, the re-initialization step can also be supplemented with the quantum-Zeno effect of the two-photon loss channel described in section VII.E. In the presence of the two-photon pump, the two-photon (or single-photon) loss channel will ``refocus" any excitations in $\mathcal{C}_\perp$ to the cat manifold.

\subsection{ Evolution during Q-Switch operation}
Once the Q-switch is turned on, the single-photon exchange coupling between the PCC and the readout cavity in the rotating frame is given by the Hamiltonian $\hat{H}_\mathrm{Q}=g(\hta^\dag\hta_\mathrm{r}+\hta\hta^\dag_\mathrm{r})$. For small $g$ this interaction can be re-written as 
\begin{align}
\hat{H}_\mathrm{Q}&=g\beta\left(\frac{p+p^{-1}}{2}\right)(\hta_\mathrm{r}+\hta^\dag_\mathrm{r})\hat{\tilde{\sigma}}_\mathrm{x}\nonumber\\
&-ig\beta\left(\frac{p-p^{-1}}{2}\right)(\hta_\mathrm{r}-\hta^\dag_\mathrm{r})\hat{\tilde{\sigma}}_\mathrm{y}.
\label{read}
\end{align}
Ignoring the term $\propto \hat{\tilde{\sigma}}_\mathrm{y}$, which becomes negligibly small even for moderately large $\beta$, the result of the Q-Switch is to displace the readout cavity conditioned on the state $(\catp\pm\catm)/\sqrt{2}$ of the PCC. If the state of the PCC is $(\catp\pm\catm)/\sqrt{2}$ then the field in the readout evolves as 
\be
\langle\hta_\mathrm{r}\rangle=\mp \frac{2 gi\beta}{\kappa_\mathrm{r}}(1-e^{-\kappa_\mathrm{r} t/2}).
\label{ideal}
\ee
To numerically confirm the above analysis we simulate the master equation $\dot{\hat{\rho}}=-i[\hat{H},\hat{\rho}]+\kappa_\mathrm{r}\mathcal{D}[\hta_\mathrm{r}]\rho$ where $\hat{H}=\hat{H}_\mathrm{pcc}+g(\hta^\dag\hta_\mathrm{r}+\hta\hta^\dag_\mathrm{r})$. The readout resonator is initialized in vacuum and the state of the PCC is along the $+X$ axis, $\ket{+}=(\catpa+\catma)/2$ or the $-X$ axis $\ket{-}=(\catpa-\catma)/2$. Figure~\ref{read_fig}(a,b) shows the evolution of $\langle\hta_\mathrm{r}\rangle$ for $\beta=1,\sqrt{2},2$, $P=K\beta^2$, $\kappa_\mathrm{r}=K/20$ and $g=\kappa_\mathrm{r}/(2\beta)$. The value for $g$ is chosen so that $\langle\hta_\mathrm{r}\rangle_\mathrm{max}=2g\beta/\kappa_\mathrm{r}=1$ and $R_\mathrm{ideal}$ remains the same irrespective of $\beta$. The time-scale for the evolution is taken to be $10/\kappa_\mathrm{r}$. For $\beta=\sqrt{2},2$ numerically obtained $\langle\hta_\mathrm{r}\rangle$ is in excellent agreement with Eq.~\eqref{ideal}. This shows that indeed for large $\beta$ the affect of the second term in Eq.~\eqref{read} is negligible. For small $\beta$ the second term in Eq.~\eqref{read} is not negligible. Therefore coupling with the readout-resonator rotates the PCC state around the $Y$-axis which leads to decrease in $\textrm{Im}[\langle\hta_\mathrm{r}\rangle]$ (Figure~\ref{read_fig}(a,b)).


\bibliographystyle{apsrev4_2}
 \bibliography{paper_arxiv.bbl}{}
 \section{Acknowledgements}
 Research was sponsored by the National Science Foundation (NSF) and by th Army Research Office (ARO), and was accomplished under Grant Numbers W911NF-18-1-0212 and W911NF-14-1-0011. The views and conclusions contained in this document are those of the authors and should not be interpreted as representing the official policies, either expressed or implied, of the Army Research Office (ARO), or the U.S. Government. The U.S. Government is authorized to reproduce and distribute reprints for Government purposes notwithstanding any copyright notation herein. S.P. would like to thank Steven Touzard and Yaxing Zhang of Yale University and Pavithran Iyer and Anirudh Krishna of University of Sherbrooke for useful discussions. 
\end{document}